\documentclass[aps,prd,nofootinbib,superscriptaddress,preprintnumbers,11pt, eqsecnum]{revtex4-1}

\usepackage[dvips]{graphicx}
\usepackage{epsf,subfigure}
\usepackage{amsfonts,amsmath}
\usepackage{relsize}
\usepackage{multirow}
\usepackage{array}
\usepackage{verbatim}
\usepackage[colorinlistoftodos]{todonotes}

\usepackage{amssymb}
\usepackage{amsmath}
\usepackage{epsfig}
\usepackage{hyperref}
\usepackage{breakurl}
\usepackage{bm}
\usepackage{color}
\usepackage{tikz}

\usepackage{amsmath}
\usetikzlibrary{arrows.meta,shapes.geometric}

\newcommand{\Hcal}{\mathcal H}
\newcommand{\Gcal}{\mathcal G}
\newcommand{\Lcal}{\mathcal L}
\newcommand{\Mcal}{\mathcal M}
\newcommand{\Tcal}{\mathcal T}
\newcommand{\Ocal}{\mathcal O}

\newcommand{\Ncal}{\mathcal N}

\newcommand{\tr}{\operatorname{Tr}}

\def\sandpm#1.#2.#3{%
  \left\langle\smash{#1}{\vphantom1}^{+}\right|{#2}%
  \left|\smash{#3}{\vphantom1}^{-}\right\rangle}
\def\spa#1.#2{\left\langle#1\,#2\right\rangle}
\def\spb#1.#2{\left[#1\,#2\right]}
\def\spab#1.#2.#3{\langle\mskip-1mu{#1}^- 
                  | #2 | {#3}^-\mskip-1mu\rangle}
\def\spba#1.#2.#3{\langle\mskip-1mu{#1}^+ 
                  | #2 | {#3}^+\mskip-1mu\rangle}
\def\spbb#1.#2.#3{\langle\mskip-1mu{#1}^+ 
                  | #2 | {#3}^-\mskip-1mu\rangle}

\newbox\charbox
\newbox\slabox
\def\s#1{{      
        \setbox\charbox=\hbox{$#1$}
        \setbox\slabox=\hbox{$/$}
        \dimen\charbox=\ht\slabox
        \advance\dimen\charbox by -\dp\slabox
        \advance\dimen\charbox by -\ht\charbox
        \advance\dimen\charbox by \dp\charbox
        \divide\dimen\charbox by 2
        \raise-\dimen\charbox\hbox to \wd\charbox{\hss/\hss}
        \llap{$#1$}
}}

\def\eps{\epsilon}
\makeatletter
\def\frontmatter@preabstractspace{\dimexpr1.5\baselineskip+1cm\relax}
\makeatother

\def\fig#1{Fig.~\ref{#1}}
\def\eqn#1{Eq.~\eqref{#1}}
\def\sect#1{Sec.~\ref{#1}}

\def\GRmatterPlus{{GR$_{\Psi}$}}

\input{tikz_figures.tex}

\begin{document}

\title{Two-Loop $n$-Point Identical Helicity Amplitude\\ 
     for Gravity Minimally Coupled to a Chiral Fermion}

\author{Zvi Bern}
\affiliation{
    Mani L. Bhaumik Institute for Theoretical Physics,
    University of California at Los Angeles,
    Los Angeles, CA 90095, USA}

\author{Calvin (Linhan) Chen}
\affiliation{
    Mani L. Bhaumik Institute for Theoretical Physics,
    University of California at Los Angeles,
    Los Angeles, CA 90095, USA}

\author{Enrico Herrmann}
\affiliation{
	Mani L. Bhaumik Institute for Theoretical Physics,
	University of California at Los Angeles,
	Los Angeles, CA 90095, USA}

\author{Anthony Morales}
\affiliation{Department of Physics, University of Washington, Seattle, WA 98195, USA}

\author{Radu Roiban}
\affiliation{Institute for Gravitation and the Cosmos and 
    Institute for Computational and Data Sciences\\
    Pennsylvania State University,
    University Park, PA 16802, USA}
\affiliation{Institute for Theoretical Studies, ETH Zurich, 8092 Zurich, Switzerland}

\author{Michael~S.~Ruf}
\affiliation{Stanford Linear Accelerator Center, 
    Menlo Park, United States}


\begin{abstract}

Motivated by recent developments in the two-loop identical-helicity sector of gauge theory when the one-loop self-dual integrability anomaly of Bardeen and Cangemi, or the equivalent twistor-space version of Costello and Paquette, vanishes, we compute the two-loop four-point identical-helicity amplitude in gravity coupled to a single chiral fermion, for which the corresponding self-dual anomaly also vanishes. We show that the infrared-subtracted amplitude is purely rational and proportional to the identical-helicity tree amplitude of pure gravity with a single insertion of the operator $R^3$. Using this connection, we obtain a closed-form expression for the two-loop $n$-point identical-helicity amplitude in gravity minimally coupled to a fermion.  We do so by first obtaining the corresponding identical-helicity gauge-theory tree amplitude with a single $F^3$ insertion using a chiral bootstrap. We then convert these $n$-point gauge theory amplitudes using Kawai--Lewellen--Tye to obtain gravity amplitudes with an $R^3$ insertion, which are necessarily equal to the desired two-loop identical helicity amplitudes of gravity coupled to a chiral fermion. As a nontrivial check, we explicitly verify crossing symmetry and the expected soft limits of our expressions through 13 points.
\end{abstract}

\maketitle

\tableofcontents

\newpage 

\section{Introduction} 

Identical-helicity amplitudes in gauge and gravity theories are remarkably simple, making them important theoretical laboratories for uncovering hidden structures, testing computational methods, and probing relations between gauge theory and gravity (see e.g. Refs.~\cite{vanNieuwenhuizen:1976vb, Mangano:1987xk, Mahlon:1993si, Bern:1993qk, Bern:1998xc, Bern:1998sv, Bern:2000dn, Broedel:2012rc, Mogull:2015adi, Dunbar:2016aux, Dunbar:2019fcq, Badger:2019djh, Dunbar:2020wdh, Dalgleish:2020mof, Chattopadhyay:2021udc, Costello:2023vyy, Dixon:2024mzh, Dalgleish:2024sey}). For minimally coupled Yang--Mills and Einstein--Hilbert theories, identical-helicity amplitudes with external gluons or gravitons vanish at tree level~\cite{Grisaru:1976vm, Grisaru:1977px, Mangano:1990by, Dixon:1996wi}, while their nonvanishing one-loop counterparts are rational functions~\cite{Bern:1993qk, Bern:1998xc} rather than polylogarithmic ones.
Remarkably, at one loop, the all-plus helicity amplitudes of Yang-Mills theory and Einstein gravity coincide with those of their self-dual counterparts~\cite{Bern:1993qk, Bern:1998sv, Bardeen:1995gk, Cangemi:1996rx, Chalmers:1996rq, Doran:2023cmj}.   The parity-conjugate all-minus helicity amplitudes are similarly related to the anti-self-dual theories.   This correspondence links the simplicity of these amplitudes to the structure of the self-dual theories. 
Changing the minimally coupled massless matter content affects only an overall state-counting coefficient in one-loop all-plus gauge-theory primitive and gravity amplitudes with external gluons or gravitons; the full color-dressed gauge-theory amplitude also depends on the matter’s color representation~\cite{Bern:1993qk, Bern:1994zx, Bern:1998sv}.

Connections of these amplitudes to celestial and twisted holography~\cite{Cheung:2016iub, Pate:2019lpp, Fan:2019emx, Himwich:2021dau, Crawley:2023brz, Bittleston:2024efo}, twistor space~\cite{Witten:2003nn, Mason:2009afn, Bittleston:2022nfr, Bittleston:2022jeq}, and the algebraic structures associated with soft and collinear limits~\cite{Strominger:2021mtt, Guevara:2021abz, Ball:2021tmb, Mago:2021wje, Ren:2022sws, Monteiro:2022lwm, Costello:2022upu, Ball:2023sdz, Costello:2023vyy, Bittleston:2023bzp, Fernandez:2024qnu, Morales:2025alm} have attracted renewed interest. In the proposed celestial conformal field theory (CCFT) framework, tree-level soft and collinear limits of gauge and gravity amplitudes are encoded in current algebras: the soft-gluon sector gives a color current algebra, whereas the conformally soft graviton sector contains a $w_{1+\infty}$ algebra~\cite{Pate:2019lpp, Guevara:2021abz, Strominger:2021mtt}. We reserve the term chiral algebra (CA) for the twistor-space constructions of Costello, Paquette, and Bittleston~\cite{Costello:2022wso, Costello:2022upu, Bittleston:2022jeq}.  Here we formulate the bootstrap in this CA framework. The self-dual sectors of Yang--Mills theory (SDYM) and gravity (SDGR) are classically integrable, with infinite towers of conserved currents associated with hidden symmetries~\cite{Chau:1982mj, Bardeen:1995gk, Popov:1996uu}.

At one loop, the nonvanishing identical-helicity amplitudes are tied to an anomaly in the conserved currents of the self-dual theories~\cite{Bardeen:1995gk, Cangemi:1996rx, Doran:2023cmj}. They can be interpreted as a one-loop-exact obstruction to integrability and, in the twistor-space description of self-dual gravity, to the corresponding Hamiltonian diffeomorphisms~\cite{Costello:2021bah, Bittleston:2022nfr, Bittleston:2022jeq}. In the CA framework~\cite{Costello:2022wso},
the twistor-space anomaly manifests as a failure of associativity of the one-loop operator product expansion in self-dual Yang--Mills theory~\cite{Costello:2022upu}. The corresponding obstruction to associativity of the CA in gravity was studied in Refs.~\cite{Bittleston:2022nfr, Bittleston:2022jeq}. We refer collectively to these related obstructions as ``anomalies.'' Supersymmetry provides a familiar way to cancel the self-dual anomaly, while also forcing both identical-helicity and single-minus amplitudes to vanish to all orders in perturbation theory~\cite{Grisaru:1976vm, Grisaru:1977px}. More generally, in minimally coupled massless gravity theories with balanced bosonic and fermionic degrees of freedom, both helicity sectors vanish at one loop in four dimensions~\cite{Bern:1993qk, Bern:1994zx, Bern:1998sv, Alston:2015gea}. Here we instead cancel the anomaly without supersymmetry, allowing a nonzero rational all-plus amplitude at two loops.

Recent work in gauge theory demonstrated that, after canceling the anomaly with a suitable matter content and removing universal infrared divergences, the two-loop all-plus Yang--Mills amplitudes are rational functions: Costello derived the single-trace amplitudes using the CA bootstrap~\cite{Costello:2022wso, Costello:2023vyy}, while Dixon and Morales 
computed the double-trace terms via the CA bootstrap, and also confirmed the results in the full theory at four points via amplitude methods~\cite{Dixon:2024mzh}. The full theory requires regularization and subtraction of universal infrared singularities~\cite{Weinberg:1965nx, Kunszt:1994np, Catani:1998bh}, which would otherwise leave kinematic-dependent logarithms.

The recent gauge-theory result raises a natural question: do two-loop
identical-helicity gravity amplitudes likewise become rational when the
one-loop anomaly is canceled? A motivation for exploring this
possibility is the double-copy relationship between gauge theory and
gravity~\cite{Bern:2008qj, Bern:2010ue, Bern:2019prr}. This relationship
takes a particularly direct form in the self-dual
sectors~\cite{Monteiro:2011pc, Chacon:2020fmr} and extends to their
anomalous one-loop amplitudes, which admit a double-copy
relation even after loop integration~\cite{Monteiro:2022lwm}.
This expectation is further supported by the CA of self-dual gravity~\cite{Bittleston:2022nfr, Bittleston:2022jeq}, where the one-loop anomaly obstructs associativity, and its cancellation restores a consistent algebra, which is expected to result in rational identical-helicity amplitudes at two loops.

To test this expectation, we explicitly compute the two-loop four-graviton all-plus helicity amplitude in Einstein gravity coupled to a single massless Weyl fermion. The two fermionic degrees of freedom balance the two graviton degrees of freedom, canceling the one-loop self-dual anomaly. After infrared subtraction and ultraviolet renormalization, all logarithmic terms cancel, leaving a rational amplitude proportional to the tree amplitude with a single insertion of $R^3 \equiv R^{\mu\nu}_{\alpha\beta}R^{\alpha\beta}_{\sigma\rho}R^{\sigma\rho}_{\mu\nu}$. This explicit four-point result fixes the normalization of our higher-point construction.

At higher multiplicity, the vanishing of the relevant two- and three-particle unitarity cuts in four dimensions supports the absence of branch-cut logarithms, as discussed in section~\ref{sec:chiral_algebra_bootstrap}. Regulator-suppressed contributions require careful infrared subtraction and ultraviolet renormalization, which we demonstrate explicitly at four points. Rationality alone does not establish that the full two-loop amplitude equals the $R^3$-insertion tree amplitude: this identification also requires matching factorization and excluding independent contact terms or recursion boundary contributions. We therefore distinguish the construction of the $R^3$ amplitude family from its identification with the full renormalized two-loop amplitudes.  

The identical-helicity sector of the $R^3$-insertion tree amplitude has already been studied in some detail in Ref.~\cite{Dunbar:2017dgp}, where the authors wrote down an on-shell recursion for any number of external legs, assuming that there are no unaccounted poles at infinity in their three-line shift.  Ref.~\cite{Dunbar:2017dgp} explicitly solved the recursion relation through eight points. For related work, see also the all-line shift of Ref.~\cite{Cohen:2010mi}, under which the identical-helicity $n$-point amplitude with an $R^3$ insertion leads to a valid recursion for $n>3$.

To simplify the higher-point construction, we first work in gauge theory, where color ordering organizes the collinear singularities and the chiral-algebra bootstrap directly computes local-operator form factors. KLT relations then convert these gauge-theory building blocks into gravity amplitudes, avoiding a more complicated direct gravitational bootstrap. 

Instead of a recursive definition, we obtain a closed-form expression for the all-plus $F^3$ form factor using the gauge-theory chiral-algebra bootstrap~\cite{Costello:2022wso}. The operator product expansion fixes its collinear residues, while the conformal weights constrain its behavior at infinity and remove the regular ambiguities in the reconstructions used here. Taking the momentum-conserving limit of this form factor gives the gauge-theory amplitudes with a single $F^3$ insertion, which we use in the Kawai--Lewellen--Tye~\cite{Kawai:1985xq} (KLT) double-copy relations to construct the corresponding gravity amplitudes, following Ref.~\cite{Broedel:2012rc}. The double copy includes dilaton and axion exchange in addition to the pure-graviton contribution; we account for these additional states when relating it to the amplitudes with a single $R^3$ insertion.  As independent checks of the closed-form construction, whose crossing symmetry and soft limits are not manifest, we explicitly verify crossing symmetry and leading soft limits through 13 points. We also match the expressions from Ref.~\cite{Dunbar:2017dgp} through seven points.\footnote{\linespread{0.9}\selectfont We thank David Dunbar for providing us with computer-readable files of these tree amplitudes.}

Our bootstrap starts with a direct calculation of the two-loop four-point all-plus-helicity amplitude for gravity minimally coupled to a chiral fermion. The net result is that it is proportional to the corresponding tree-level amplitudes with a single $R^3$ insertion. We evaluate the two-loop amplitude using a formal polarization-vector framework~\cite{FuturePaper}, treating all helicity configurations simultaneously and allowing fermions and scalars to circulate in the loops. The computation uses a refinement~\cite{Bern:2024vqs} of generalized unitarity~\cite{Bern:1994zx, Bern:1994cg, Bern:1997sc, Britto:2004nc, Bern:2004cz, Bern:2007ct}, tensor projection (see Refs.~\cite{Garland:2002ak, Peraro:2020sfm} for a gauge-theory discussion), integration-by-parts reduction~\cite{Chetyrkin:1981qh, Tkachov:1981wb, Laporta:2000dsw, Lange:2025fba, Guan:2024byi, FIRE7, Maierhofer:2017gsa, Smirnov:2023yhb, Smirnov:2008iw}, and canonical differential equations~\cite{Kotikov:1990kg, Bern:1993kr, Remiddi:1997ny, Gehrmann:1999as, Henn:2013pwa, Henn:2013woa}. The formal polarization framework makes it easier to control evanescent effects in dimensional regularization. Details of the calculation, and results for all helicities, will be presented in Ref.~\cite{FuturePaper}.

An important aspect of our two-loop calculation is the detailed account of the ultraviolet (UV) and infrared (IR) singularities. Controlling evanescent effects is critical for understanding the UV properties of gravity theories~\cite{Bern:2015xsa, Bern:2017puu}, especially in $\mathcal N=4$ supergravity, where evanescent contributions are intertwined with the $U(1)$ duality anomaly and may affect the interpretation of the four-loop UV divergence~\cite{Carrasco:2013ypa, Bern:2017tuc, Bern:2017rjw, Bern:2013uka}. At one loop, four-dimensional pure-gravity amplitudes are UV finite because the Gauss--Bonnet counterterm is evanescent, i.e. it vanishes on shell in four dimensions~\cite{tHooft:1974toh}. In contrast, the matter sector can have divergences.
At two loops, Goroff and Sagnotti~\cite{Goroff:1985sz, Goroff:1985th} first computed the divergence in the $\overline{\rm MS}$ scheme, with the result confirmed in multiple computations~\cite{vandeVen:1991gw, Bern:2015xsa, Bern:2017puu, Dunbar:2017qxb, Abreu:2020lyk}, including the one reported in this paper. To obtain the correct result, one must subtract the UV subdivergences in dimensional regularization, even though the one-loop amplitude is finite in four dimensions. These subdivergences can be removed by introducing the Gauss--Bonnet counterterm or by direct subtraction of the subdivergences.

A surprising feature of the two-loop divergence is that it is scheme dependent and \emph{not} correlated with the coefficient of renormalization-scale dependence. In contrast, the coefficient of $\log\mu^2$, where $\mu$ is the renormalization scale, is scheme independent and can be computed in four dimensions, without introducing a UV regulator~\cite{Bern:2017puu, Dunbar:2017qxb}, making it a scheme-independent test of the UV properties of gravity amplitudes. This coefficient vanishes precisely when the self-dual anomaly cancels, i.e. when the bosonic and fermionic degrees of freedom are balanced. This cancellation occurs in supergravity~\cite{Grisaru:1976vm, Grisaru:1977px, Bern:2017puu} and more generally in nonsupersymmetric theories with balanced states~\cite{Bern:2015xsa, Bern:2017puu, Dunbar:2017qxb, Abreu:2020lyk}. In such anomaly-free theories, the $R^3$ coefficient absorbs the remaining divergence but has no two-loop running~\cite{Bern:2017puu}.

The infrared singularities also contain a dimensional-regularization subtlety. Once the anomaly is canceled, the four-dimensional one-loop all-plus amplitude vanishes. Together with the vanishing tree amplitude, this implies that there is no two-loop IR pole. However, regulator-suppressed one-loop terms can combine with the universal IR pole to leave finite logarithmic terms. A proper IR subtraction removes these terms, as we show explicitly, leaving an amplitude that is purely rational after UV renormalization.

The remaining rational amplitude has an overall coefficient that depends on the finite $R^3$ coupling and the four-fermion contact coupling $c_{4\Psi}$, as shown in Eq.~\eqref{eq:chiral_alpha}. Changes of subtraction scheme shift the finite terms, while $c_{4\Psi}$ is an independent parameter of the Lagrangian. Anomaly cancellation removes the two-loop renormalization-scale dependence but does not fix this finite normalization. For a fixed value of $c_{4\Psi}$, a finite $R^3$ renormalization condition can set the coefficient to zero, and this condition is stable under changes of the renormalization scale at two loops. Within the all-multiplicity construction, this choice sets the entire all-plus sector to zero through two loops without supersymmetry. In a supersymmetric theory, the vanishing would instead be enforced by a supersymmetry Ward identity.

The paper is organized as follows. In section~\ref{sec:BasicReview}, we define the chiral-fermion gravitational theory and review lower-loop identical-helicity amplitudes, self-dual sectors, and anomaly cancellation. In section~\ref{sec:chiral_algebra_bootstrap}, we state the four-point seed, review the chiral-algebra bootstrap, present the all-plus helicity $F^3$ form factor and its amplitude limit, and then use the KLT relations to obtain the all-multiplicity two-loop gravity amplitudes. We also discuss deformations, associativity, and checks of this expression. Section~\ref{sec:FourPointTwoloopAmpitude} summarizes the explicit calculation of the two-loop four-point amplitude used to seed the bootstrap. This includes the bare amplitude, infrared subtraction, ultraviolet counterterms, and the renormalized result. We conclude in section~\ref{sec:conclusions}. Appendix~\ref{app:F3_derivation} gives the detailed derivation of the $F^3$ form factor.

\section{Theory and lower-loop amplitudes}
\label{sec:BasicReview}

We begin by describing Einstein gravity coupled to a chiral fermion, denoted GR$_\Psi$, and the curvature counterterms relevant to external-graviton amplitudes.
We then review the lower-loop helicity amplitudes and anomaly cancellation in the self-dual sector. These provide the input for the four-point calculation in \sect{sec:FourPointTwoloopAmpitude} and the all-multiplicity construction in \sect{sec:chiral_algebra_bootstrap}.

\subsection{Definition of chiral fermion gravitational theory}
\label{subsec:chiral_fermion_theory}

The Lagrangian defining the GR$_\Psi$ theory, i.e. Einstein's gravity coupled to a chiral fermion, is 
\begin{align}
\mathcal{L}_{{\rm GR}_\Psi} =  
\frac{2}{\kappa^2} R 
+ i \bar{\psi}_L \gamma^a  e_a^{\; \mu}  D_{\!\mu} \psi_L 
 +   c_{4\Psi}\frac{\kappa^2}{32} \left(\bar{\psi}_L\gamma^{abc} \psi_L \right)^2 \ .
\label{LagrangianChiral}
\end{align}
Newton's constant $G$ is related to the coupling $\kappa$ via $\kappa^2 = 32 \pi G$, $e_a^{\;\mu}$ is the vielbein, $D_{\!\mu}$ is the tangent space covariant derivative and the graviton field is defined as $g_{\mu\nu} = \eta_{\mu\nu}+\kappa h_{\mu\nu}$. 
While the minimal coupling of a fermion with gravity (second term in Eq.~\eqref{LagrangianChiral}) is universal, non-minimal couplings are possible. 
The third term, tagged by $c_{4\Psi}$, leads to contributions of the same dimension as the other two and can be obtained (for $c_{4\Psi}=1/4$) by integrating out the spin connection in the first-order formalism~\cite{Freedman:2012zz} where $\gamma^{abc} = \gamma^{[a} \gamma^{b}\gamma^{c]}$ rather than $\gamma^a$ naturally appears. 
With the normalizations in Eq.~\eqref{LagrangianChiral}, the graviton and fermion have standard mass dimension, $[h_{\mu\nu}]=1$ and $[\psi_L]=3/2$.
We note that the chiral matter does not introduce spacetime diffeomorphism anomalies in $D=4$~\cite{Alvarez-Gaume:1983ihn}.
The four-fermion contact interaction does not affect tree or
one-loop amplitudes with only external gravitons, but contributes
to the finite two-loop four-graviton amplitude computed~in~\sect{sec:FourPointTwoloopAmpitude}.

Renormalization of power-counting non-renormalizable theories requires introducing higher-dimension operators as counterterms. 
The curvature counterterms relevant to external-graviton amplitudes through two loops are
\begin{align}
\Lcal_{{\rm CT}}=  
 \bigg[ & 
    L^{(1)} (\epsilon) \frac{c_{{\rm GB}}}{\epsilon}\,  E_4
    + \left(\frac{\kappa}{2}\right)^2 L^{(2)} (\epsilon) \frac{c_{{\rm R}^3}}{\epsilon} R^{\alpha \beta}_{\!\hphantom{\alpha \beta}\gamma \delta}\,
    R^{\gamma \delta}_{\!\hphantom{\gamma \delta}\rho \sigma}\,
    R^{\rho \sigma}_{\!\hphantom{\rho \sigma}\alpha \beta} 
 \bigg] \,,
 \label{LagrangianCT}
\end{align}
where $E_4 = R^2 -  4 R_{\mu \nu}R^{\mu\nu} + R_{\mu \nu \rho \sigma} R^{\mu \nu \rho \sigma}$ is the Gauss-Bonnet combination which is on-shell equivalent to a Riemann-square term. In four dimensions, the density $\sqrt{-g}\,E_4$ is locally a total derivative. The one- and two-loop normalization factors are
\begin{equation}
L^{(1)}(\epsilon) = \frac{e^{-\gamma_{\rm E}\epsilon}}{(4\pi)^{2-\epsilon}}\,,
\qquad
L^{(2)}(\epsilon) = \frac{e^{-2\gamma_{\rm E}\epsilon}}{(4\pi)^{4-2\epsilon}}\,. 
\label{LNormalization}
\end{equation}
Here we use dimensional regularization~\cite{tHooft:1972tcz} with spacetime dimension $D=4-2\epsilon$ and $\gamma_{\rm E}$ is the Euler--Mascheroni constant. These factors include all the finite terms appropriate to the $\overline{\rm MS}$ scheme.
In $D$ dimensions, there is a second independent $R^3$ counterterm, but it will not play a role for two-loop amplitudes with four-dimensional external legs~\cite{Goroff:1985sz, Goroff:1985th} where the two $R^3$ operators are equivalent. 
Additional counterterms are necessary in Eq.~(\ref{LagrangianCT}) to renormalize amplitudes with external matter, as divergences already appear at one loop~\cite{tHooft:1974toh, Deser:1974xq}.  
We note, however, that the  coefficient $c_{4\Psi}$ is dimensionless ($[\psi_L]=3/2$ with the normalization in Eq.~\eqref{LagrangianChiral}) so, on dimensional grounds, the last term in Eq.~\eqref{LagrangianChiral}  cannot be renormalized to any order in perturbation theory~\cite{Barvinsky:1981rw} in dimensional regularization.\footnote{\linespread{0.9}\selectfont For example, one-loop Feynman diagrams imply that a four-fermion term comes with a factor of $\kappa^4$ and therefore must also contain two derivatives for its dimension to be 4, which makes it different from the last term in Eq.~\eqref{LagrangianChiral}.  With a dimensionful regulator $\Lambda$, the two additional units can be supplied by e.g. $\Lambda^2$ and thus the four-fermion term may be renormalized in such a scheme.}  


\subsection{Identical-helicity amplitudes at lower loop orders}\label{subsec:review_tree_one_loop_all_plus}

The tree-level identical-helicity amplitudes of minimally-coupled gauge and gravity amplitudes all vanish for $n\ge 4$, e.g.\footnote{\linespread{0.9}\selectfont The all-minus amplitudes follow directly from parity conjugation and are often left implicit in this work, where we synonymously use ``identical helicity amplitudes'' and ``all-plus amplitudes''.}
\begin{equation}
\Mcal^{(0)}_{n}(1^+, 2^+, 3^+, \ldots, n^+) =  0\,.
\label{eq:FourPointTreeVanish} 
\end{equation}
For generic kinematics, this identity can be proved either by an appropriate choice of reference spinors for the external polarization vectors~\cite{Dixon:1996wi, Elvang:2013cua}, or by choosing a suitable gauge for the polarization vectors, or by applying supersymmetry Ward identities~\cite{Grisaru:1977px}; the latter also applies to pure-graviton tree amplitudes because embedding Einstein gravity into supergravity leaves these amplitudes unchanged. Similar considerations also hold for the gauge-theory case and for amplitudes where a single helicity is flipped from plus to minus. The vanishing in Eq.~\eqref{eq:FourPointTreeVanish} refers to Einstein amplitudes without higher-curvature insertions. A single $R^3$ insertion generates the all-plus tree amplitudes used below.

At one loop, all-plus helicity amplitudes are generally nonzero but remain remarkably simple~\cite{Bern:1993qk, Bern:1998xc}. In any theory of massless particles with spin $s<2$ minimally coupled to gravity, the one-loop four-graviton amplitude with all positive-helicity external gravitons is
\begin{align}
\Mcal^{(1)}_{4}(1^+,2^+,3^+,4^+)
= -i\,
\frac{1}{(4\pi)^2}\left(\frac{\kappa}{2}\right)^4
\Tcal^2\,\frac{N_b-N_{\!f}}{240}
(s^2{+}t^2{+}u^2)
+\Ocal(\epsilon)\,,
\label{eq:FourPointOneLoopAmplGen}
\end{align}
where $N_b$ and $N_{\!f}$ denote the numbers of bosonic and fermionic states circulating in the loop, respectively.
All external momenta are outgoing and massless, with $p_i^2=0$
and $\sum_{i=1}^4 p_i=0$. We write
$p_i^{\dot\alpha\alpha}=\tilde\lambda_i^{\dot\alpha}\lambda_i^\alpha$
and use $\langle ij\rangle=\epsilon_{\alpha\beta}
\lambda_i^\alpha\lambda_j^\beta$ and
$[ij]=-\epsilon_{\dot\alpha\dot\beta}
\tilde\lambda_i^{\dot\alpha}\tilde\lambda_j^{\dot\beta}$,
where the $\epsilon$ tensors are the $SU(2)_L$ and $SU(2)_R$-invariant two-index tensors, numerically equal to the two-dimensional Levi-Civita tensor.
The Mandelstam invariants are
$s_{ij}=(p_i+p_j)^2=\langle ij\rangle[ji]$;
at four points, $s=s_{12}$, $t=s_{23}$, and $u=s_{13}$,
with $s+t+u=0$ by momentum conservation.
The factor,
\begin{equation}
\Tcal \equiv
\frac{[12][34]}{\langle12\rangle\langle34\rangle}\,,
\label{eq:Tfactor}
\end{equation}
carries the helicity weight and is completely Bose symmetric.

Equation~\eqref{eq:FourPointOneLoopAmplGen} shows that, apart from the overall factor $N_b-N_{\!f}$, the four-dimensional one-loop four-point identical-helicity amplitude is independent of the particle content. The cancellation when $N_b=N_{\!f}$ does not require supersymmetry. A graviton contributes $N_b=2$ and a chiral fermion contributes $N_{\!f}=2$, so the $\Ocal(\epsilon^0)$ all-plus amplitude vanishes in GR$_\Psi$.
This reflects the result of Refs.~\cite{Bern:1993qk, Bern:1994zx,Bern:1998sv} that bosonic and fermionic contributions to these one-loop amplitudes are equal up to a sign, derived there by applying supersymmetry Ward identities to linear combinations of supermultiplet contributions.
In supersymmetric theories, supersymmetry Ward identities enforce the vanishing of these helicity sectors to all orders in perturbation theory~\cite{Grisaru:1976vm, Grisaru:1977px, Parke:1985pn, Kunszt:1985mg}.

The one-loop amplitude is also both infrared and ultraviolet finite: the one-loop infrared singularities are proportional to the vanishing all-plus tree amplitude, while the possible one-loop Gauss--Bonnet UV counterterm is evanescent~\cite{tHooft:1974toh}. All-plus one-loop amplitudes with an arbitrary number of external legs were obtained for theories with minimally coupled massless matter in Refs.~\cite{Bern:1998xc, Bern:1998sv}; analogous amplitudes with massive particles can be analyzed similarly~\cite{Bern:1995db}.

While the four-dimensional $\Ocal(\epsilon^0)$ part of the one-loop amplitude vanishes, $\Ocal(\epsilon)$ and higher terms remain nonzero in the 't~Hooft--Veltman (HV) regularization scheme used here, because the bosonic and fermionic degrees of freedom are continued differently away from four dimensions. As we will see in \sect{sec:FourPointTwoloopAmpitude}, these $\Ocal(\epsilon)$ terms of the four-point one-loop amplitude enter the two-loop infrared subtraction; see Eqs.~\eqref{HardPart} and~\eqref{eq:2loop_IR_subtraction}. 
We provide the required $\Ocal(\epsilon)$ terms in the ancillary file~\cite{AncillaryFile} and discuss them further in \sect{subsec:review_amp_ir_divs}.

\subsection{Self-dual sectors and anomaly cancellation}
\label{subsec:review_chiral_algebra}

Both Yang-Mills and Einstein gravity have first-order formulations as deformations of their self-dual sectors. The self-dual sector of gauge theory is described by the $g\to0$ limit of the Chalmers-Siegel (CS) action~\cite{Chalmers:1996rq}
\begin{equation}
\label{CSaction}
S_\mathrm{CS} = \int d^4x\,\tr\!\left(BF_-+\frac{g^2}{2}B^2\right)
\hspace{0.5cm}
\overset{g\to0}{\longrightarrow} 
\hspace{0.5cm}
S_\mathrm{SDYM} = \int d^4x\,\tr(BF_-)
\,,
\end{equation}
where $B$ is an adjoint-valued anti-self-dual (ASD) two-form field and $F_-$ is the anti-self-dual component of the field strength $F$. The self-dual and anti-self-dual components of $F$ are defined by
\begin{equation}
F_\pm = \frac{1}{2}(F\pm\widetilde F)
\,,
\end{equation}
where the dual of $F$ is defined in components through the Minkowski Levi-Civita symbol $\epsilon_{\mu\nu\rho\sigma}$ as
\begin{equation}
    \widetilde F_{\mu\nu} = \frac{i}{2}\epsilon_{\mu\nu\rho\sigma}F^{\rho\sigma}
\end{equation}
where $\epsilon_{0123}=+1$. The Chalmers-Siegel action imposes $B=-g^{-2}F_-$ on shell and, upon integrating out $B$, reduces to the Yang-Mills action plus a topological term. In the self-dual limit, the equations of motion impose the self-duality condition $F_-=0$, as well as $D_\mu B^{\mu\nu}=0$. The linearized equations of motion show that the gauge field $A$ and the ASD two-form $B$ describe gluons of positive and negative helicities, respectively. So, we see that SDYM only has a $(-++)$ vertex and $\tr(B^2)$ restores the opposite helicity structure $(--+)$.

Analogously, SDGR is given by the $\kappa\to0$ limit of the chiral Plebanski action for pure gravity with no cosmological constant, after rescaling the chiral spin connection, as \cite{Plebanski:1977zz,Abou-Zeid:2005zfo}
\begin{equation}
\label{plebanski}
S_{\mathrm{Plebanski}}
=\frac{1}{2}\int e^{\alpha}{}_{\dot\alpha}\wedge e^{\dot\alpha\beta}\wedge
\left(d\Gamma_{\alpha\beta}
-\frac{\kappa^2}{8}\Gamma_{\alpha\gamma}\wedge
\Gamma^\gamma{}_\beta\right)
\hspace{0.25cm}
\overset{\kappa\to0}{\longrightarrow}
\hspace{0.25cm}
S_{\mathrm{SDGR}}
=\frac{1}{2}\int\Gamma_{\alpha\beta}\wedge d\left(e^{\alpha}{}_{\dot\alpha}\wedge e^{\dot\alpha\beta}\right)
\,,
\end{equation}
where we used integration by parts to move the derivative off of $\Gamma_{\alpha\beta}$. Here, the $e^{\dot\alpha\alpha}$ are vierbein written with $SL(2,\mathbb{C})_\pm$ spinor indices, as opposed to a single Lorentz index. The $\Gamma_{\alpha\beta}$ are one-form fields that play the analogous role to the ASD two-form $B$ of SDYM: the Plebanski action imposes the on-shell condition that $\tfrac{\kappa^2}{8}\Gamma_{\alpha\beta}$ equals the ASD spin connection determined by the metric, whereas $\Gamma_{\alpha\beta}$ acts as a Lagrange multiplier in the self-dual limit enforcing $d(e^{\alpha}{}_{\dot\alpha}\wedge e^{\dot\alpha\beta})=0$. The closure of $e^{\alpha}{}_{\dot\alpha}\wedge e^{\dot\alpha\beta}$ implies that the ASD spin connection is flat, giving a Ricci-flat metric with self-dual Riemann curvature. These conditions imply that the vierbein $e^{\dot\alpha\alpha}$ and the one-form fields $\Gamma_{\alpha\beta}$ describe gravitons of positive and negative helicities, respectively. Thus, SDGR only has a $(-++)$ vertex, and the $(--+)$ vertex is restored upon deforming the theory by $-\tfrac{\kappa^2}{16}e^{\alpha}{}_{\dot\alpha}\wedge e^{\dot\alpha\beta}\wedge\Gamma_{\alpha\gamma}\wedge
\Gamma^\gamma{}_\beta$, which in turn generates the MHV amplitudes \cite{Mason:2009afn}.

As alluded to in the introduction, the perturbative $S$-matrices of the strict self-dual theories are extremely restricted.  Apart from the three-point tree amplitude with helicities $(-++)$, which has on shell support only on complexified kinematics, their only non-vanishing scattering amplitudes are the rational one-loop all-plus amplitudes: higher-point tree amplitudes vanish for generic kinematics, while the helicity structure of the self-dual vertices forbids connected amplitudes beyond one loop.  The same structure explains why the one-loop identical-helicity amplitudes of full Yang--Mills theory and Einstein gravity coincide with those of their self-dual counterparts,  i.e. 
\begin{equation}
\mathcal{A}^{(1)}_{n,\mathrm{YM}}(+\cdots+) = \mathcal{A}^{(1)}_{n,\mathrm{SDYM}}(+\cdots+)\,, \hskip 0.5 cm \text{and}
\hskip 0.5 cm 
\mathcal{M}^{(1)}_{n,\mathrm{GR}}(+\cdots+) = \mathcal{M}^{(1)}_{n,\mathrm{SDG}}(+\cdots+) \,.
\label{eq:full_self_dual_one_loop}
\end{equation}
Indeed, the self-dual actions retain precisely the $(-++)$ interaction required to assemble the relevant one-loop diagrams \cite{Bardeen:1995gk,Chalmers:1996rq,Cangemi:1996rx,Bern:1998xc,Boels:2013bi}.  The all-plus amplitudes of the full theories may therefore be viewed as quantum remnants of otherwise classically integrable chiral subsectors.

Classical integrability is encoded by an infinite tower of hidden
conserved currents, which become manifest under a non-local
transformation of the self-dual theories to free fields. At the quantum
level, the effective vertex (or equivalently, the Jacobian of this
transformation) makes these currents anomalous; its on-shell matrix
elements are the one-loop all-plus amplitudes. This connection was
first observed by Bardeen~\cite{Bardeen:1995gk} and further studied in
Refs.~\cite{Cangemi:1996rx,Doran:2023cmj,Chattopadhyay:2020oxe,
Monteiro:2022nqt}. The resulting one-loop-exact anomaly concerns the
hidden currents responsible for integrability~\footnote{\linespread{0.9}\selectfont
A related amplitude-level connection involves the $U(1)$ duality
anomaly of $\mathcal N=4$ supergravity~\cite{Marcus:1985yy}.
Its double-copy integrands combine $\mathcal N=4$ super-Yang--Mills
with ordinary Yang--Mills~\cite{Kawai:1985xq,Bern:2008qj,
Bern:2010ue,Bern:2019prr}. The Yang--Mills all-plus amplitudes enter
finite rational one-loop supergravity amplitudes that violate the
tree-level duality selection rules~\cite{Carrasco:2013ypa,
Bern:2017tuc,Bern:2017rjw,Bern:2019isl,Kallosh:2026roi}.
This connection concerns the amplitude manifestations of distinct
anomalies.} and is distinct from
an ordinary gauge anomaly.

The states of SDYM and SDGR can be shown to be in one-to-one correspondence with the local operators of a chiral algebra (CA) living on a Riemann sphere \cite{Costello:2022wso,Costello:2022upu,Bittleston:2022jeq}. These CAs are constructed by first uplifting the self-dual theories to twistor space $\mathbb{PT}$, then realizing the CAs 
\begin{enumerate}
    \item[1)] as the Lie algebra of infinitesimal gauge transformations on $\mathbb{PT}$ (with the removal of specific fibers) that preserve the vacuum, or 
    
    \item[2)] as the Koszul dual\footnote{\linespread{0.9}\selectfont See ref.~\cite{Paquette:2021cij} for a review of Koszul duality.} to the algebra of local operators in the theory on $\mathbb{PT}$. 
\end{enumerate}

\noindent
The twistor-space constructions exhibit a quantum obstruction closely related to the integrability anomalies of the self-dual theories in the usual formulation.
One-loop box diagrams compute the four-point all-plus amplitudes of the four-dimensional theories and expose gauge anomalies in the holomorphic theory on $\mathbb{PT}$~\cite{Costello:2021bah, Bittleston:2022nfr}. 
In gravity, the anomaly of the twistor-space theory can be interpreted as an anomaly in Hamiltonian diffeomorphisms~\cite{Bittleston:2022nfr, Bittleston:2022jeq}. The one-loop correction associated with the all-plus splitting function obstructs associativity of the CA~\cite{Costello:2022upu,Bittleston:2022jeq}. 

To obtain an associative chiral algebra from a twistor theory, we cancel the anomaly by introducing additional states with local interactions on $\mathbb{PT}$~\cite{Costello:2021bah,Costello:2022wso, Bittleston:2022nfr}. The coefficient of the gauge anomaly of the twistor formulation of SDGR is proportional to the boson--fermion state imbalance~\cite{Bittleston:2022nfr, Bittleston:2022jeq},
\begin{equation}
\mathcal A\propto N_b-N_{\!f}\,,
\label{eq:twistor_anomaly}
\end{equation}
where $N_b$ and $N_f$ are the number of bosonic and fermionic degrees of freedom, respectively. 

In a supersymmetric self-dual theory $N_b=N_f$ so the anomaly is absent. Supersymmetry however is not the only means to remove this anomaly. Indeed, the self-dual truncation of the full GR$_\Psi$ theory defined in~Eq.~\eqref{LagrangianChiral} is anomaly-free: its two graviton states
are balanced by the two states of a single Weyl fermion.\footnote{\linespread{0.9}\selectfont  Alternatively, a Green--Schwarz mechanism~\cite{Green:1984sg} allows $N_b-N_f\ne0$ by introducing a scalar with a fourth-order kinetic term coupled to $R^\mu_{\;\nu}\wedge R^\nu_{\;\mu}$~\cite{Costello:2021bah, Bittleston:2022nfr}. This construction represents the one-loop all-plus amplitude as a tree-level amplitude~\cite{Costello:2022wso, Bittleston:2022jeq}. Analogous gauge-theory constructions use suitable anomaly-free matter or a Green--Schwarz mechanism~\cite{Costello:2022upu}.} 
The hidden-current anomaly and the twistor-space obstruction share the $N_b-N_{\!f}$ dependence relevant to our cancellation mechanism. This common dependence is reflected in the vanishing of the one-loop amplitude for $N_b=N_{\!f}$ in \eqn{eq:FourPointOneLoopAmplGen}.

The anomaly-free gauge theories discussed in the introduction provide examples of rational two-loop all-plus amplitudes after infrared subtraction~\cite{Costello:2023vyy, Dixon:2024mzh}. In \sect{sec:FourPointTwoloopAmpitude}, we find an analogous simplification in gravity: after ultraviolet renormalization and infrared subtraction, the two-loop four-graviton all-plus amplitude in GR$_\Psi$ is rational and proportional to the tree-level four-graviton amplitude with a single insertion of the $R^3$ operator. This result concerns the full GR$_\Psi$ theory. 
The lower-loop amplitudes reviewed above supply the input for the required infrared and ultraviolet subtractions. 
We use this renormalized, infrared-subtracted four-point gravity amplitude to seed a bootstrap construction of the identical-helicity two-loop graviton amplitudes for arbitrary multiplicity. To this end, in \sect{sec:chiral_algebra_bootstrap} we use the gauge-theory chiral algebra to construct the $F^3$ form factor and take its amplitude limit\footnote{The amplitude limit of the $F^3$ form factor amounts to setting to 0 the operator momentum while preserving momentum conservation.}. Then, using the close relation between the $F^3$ and $R^3$ operators~\cite{Broedel:2012rc}, we use the KLT relations to obtain the corresponding all-multiplicity gravity expression, normalized by this four-point seed.

\section{Chiral algebra bootstrap, KLT, and two-loop $n$-point all-plus amplitudes}
\label{sec:chiral_algebra_bootstrap}

\subsection{Bootstrap overview}

In this section we describe an $n$-point bootstrap of the two-loop all-plus helicity amplitude. The key input is, on the one hand, the explicit renormalized and IR-subtracted four-point two-loop all-plus amplitudes of GR$_\Psi$ which are rational and serve as a seed for our bootstrap, 
and on the other hand, an argument that the all-multiplicity counterparts of these amplitudes are also rational. 

We compute the renormalized two-loop four-graviton all-plus amplitude in \sect{sec:FourPointTwoloopAmpitude}, and find
\begin{equation}
\Mcal^{(2),\rm ren}_{4,\rm chiral}
= 10\,i\,\alpha\left(\frac{\kappa}{2}\right)^2\Tcal^2\,stu\,,
\label{eq:AmplChiralPreview}
\end{equation}
where $\alpha$ is an overall constant depending on the renormalized Lagrangian parameters appearing in Eqs. \eqref{LagrangianChiral} and~\eqref{LagrangianCT}, as described in \sect{sec:FourPointTwoloopAmpitude}. This amplitude matches the tree amplitude with a single $R^3$ insertion, with $\alpha$ normalized as in Ref.~\cite{Dunbar:2017dgp}, whose $n$-point amplitudes have the overall factor $i(\kappa/2)^{n-2}$ stripped off. This four-point match motivates constructing the higher-point two-loop amplitudes from tree amplitudes with a single $R^3$ insertion.
Higher-point amplitudes with an $R^3$ insertion have been studied using on-shell recursion~\cite{Cohen:2010mi, Dunbar:2017dgp}, with explicit results and checks through seven points in Ref.~\cite{Dunbar:2017dgp}.

The absence of logarithms at higher points can be understood from unitarity cuts evaluated in four dimensions. Ordinary channel discontinuities at two loops involve two- or three-particle cuts. For generic channels, with at least two external gravitons on each side of the cut, a two-particle cut is a product of a tree and a one-loop amplitude. With graviton cut states, the tree amplitude is nonzero only when both cut states are negative-helicity gravitons, which then forces the one-loop amplitude on the other side of the cut to be all-plus; this amplitude vanishes when $N_b=N_f$, canceling the anomaly. Tree amplitudes containing a fermion pair and otherwise only positive-helicity gravitons also vanish, eliminating fermion two-particle cuts.

Three-particle cuts instead contain two trees. With three cut gravitons, crossing supplies only three negative helicities across the two trees, while two nonzero graviton trees require at least four. If a fermion pair crosses the cut, one tree contains that pair and otherwise only positive-helicity gravitons and vanishes by the same identity. These arguments hold independently of the external multiplicity, showing that logarithms should not appear.\footnote{\linespread{0.9}\selectfont In the presence of infrared and ultraviolet divergences,  the argument is more subtle because $O(\epsilon)$ terms can interfere with $1/\epsilon$ poles, but the basic idea is the same; in \sect{sec:FourPointTwoloopAmpitude} we show in detail the cancellations at four points.}

Once the renormalized, infrared-subtracted all-plus amplitudes are known to be rational, their higher-point bootstrap is identical to that of a single $R^3$ insertion, since standard factorization accounts for all their poles. For $n\geq5$, factorization contributions with a tree and two-loop amplitude vanish unless the tree factor is the Einstein three-point $(++-)$ amplitude. The factorization contributions with one loop on both sides of the factorization pole also vanish because the relevant one-loop helicity amplitudes cancel when $N_b =N_f$. Thus the surviving residues are determined by the Einstein three-point amplitude and lower-point two-loop all-plus amplitudes, exactly as for the $R^3$ tree family. Matching the four-point amplitudes therefore fixes all higher-point residues inductively. Any remaining difference would be a local polynomial. After stripping the gravitational coupling, this polynomial must have momentum degree six, whereas little-group covariance requires degree at least $2n$ for an all-plus $n$-graviton polynomial. No such ambiguity terms exist for $n\geq5$ uniquely fixing the bootstrap. 

In this section we obtain a closed-form formula for arbitrary multiplicity two-loop identical-helicity amplitudes. The observation of Broedel and Dixon~\cite{Broedel:2012rc} relates all-plus tree amplitudes with a single $R^3$ insertion to KLT double copies of the corresponding $F^3$-insertion amplitudes~\cite{Kawai:1985xq}. The comparison with a pure $R^3$ insertion must account for the additional states produced by the double copy. Building on~\sect{subsec:review_chiral_algebra}, we first obtain the all-plus $F^3$ form factor using the gauge-theory chiral algebra bootstrap, then take its amplitude limit and apply the closed-form KLT relations~\cite{Bern:1998sv} in the double-copy construction~\cite{Bern:2008qj,Bern:2010ue,Bern:2019prr}.
Identifying this tree-amplitude family with the full renormalized two-loop amplitudes also requires a verification of its factorization properties and detailed control of any independent contact terms or recursion boundary contributions. The form-factor derivation is given in Appendix~\ref{app:F3_derivation}.


\subsection{Chiral-algebra bootstrap framework}
\label{subsec:cab_framework}

We review the chiral algebra bootstrap (CAB) for gauge theory, which is used in Appendix~\ref{app:F3_derivation} to derive the $n$-point tree amplitudes with one insertion of the $F^3$ operator, included below in Eqs.~\eqref{eq:F3}--\eqref{eq:T6}. The CAB has been used to compute one- and two-loop form factors in non-supersymmetric self-dual gauge theories with special matter representations \cite{Costello:2022upu, Costello:2023vyy, Dixon:2024mzh}, as well as in $\mathcal{N}=4$ self-dual SYM \cite{Charanya:2026pnh}. Of particular interest are form factors with insertions of the operator $\tr(B^2)$, as they recover non-self-dual gauge theory amplitudes in the zero-operator-momenta limit. For a single $\tr(B^2)$ insertion, the form factors give the MHV, single-minus, and all-plus amplitudes at tree level, one loop, and two loops, respectively, and they have been computed with the CAB in Refs.~\cite{Costello:2022upu, Costello:2023vyy, Dixon:2024mzh}.
These amplitudes include contributions from matter content that cancels both the integrability anomaly of the self-dual theories and the gauge anomaly in the twistor space formulation of self-dual theories.  
By examining the color structure of amplitudes in various such theories, two-loop QCD amplitudes with an arbitrary number of fundamental fermions can be recovered from a linear combination of anomaly-free self-dual form factors and one-loop amplitudes~\cite{Morales:2025alm}, showing that the CAB might be useful for computing observables in physically relevant theories.

Let $J^a(\tilde\lambda,z)$ and $\tilde J^a(\tilde\lambda,z)$ represent momentum eigenstates of positive- and negative-helicity gluons, respectively, with massless momentum $p^{\dot\alpha\alpha}=\tilde\lambda^{\dot\alpha}\lambda^\alpha$ and the choice $\lambda^\alpha = (z,1)$. With this choice, it is understood that $\tilde\lambda$ depends on the energy of the state, while $z$ does not. 
Thus, the soft expansion is an expansion in the spinors $\tilde\lambda$, which may be thought of as absorbing the factor of the energy $\omega$ used for this expansion in Refs.~\cite{Costello:2022wso,Costello:2022upu}.
%
The eigenstates can then be expanded into energy modes as
\begin{equation}
J^a(\tilde\lambda,z) = \sum_{k\geq 0}J^a[k](\tilde\lambda,z) \, ,
\label{eq:Jexpansion}
\end{equation}
where each mode is assumed to scale as $(\tilde\lambda)^k$, so it may be further written as
\begin{equation}
    \label{eq:modeex}
    J^a[k](\tilde\lambda,z) = \sum_{\substack{r_1,r_2\geq0 \\ r_1+r_2=k}}\frac{(\tilde\lambda^{\dot1})^{r_1}(\tilde\lambda^{\dot2})^{r_2}}{r_1!r_2!}
    J^a[r_1,r_2](z)
    \, .
\end{equation}
Analogous expansions hold for $\tilde{J}$ and for all other operators depending on $z$ and ${\tilde\lambda}$. 
We will often suppress the dependence on $\tilde\lambda$ in the modes and simply write $J^a[k](z)$ in place of $J^a[k](\tilde\lambda,z)$. 

The two-dimensional chiral algebra of the self-dual gauge theory can be defined classically as the collection of generators 
$J^a[r_1,r_2](z)$ and $\tilde{J}^a[r_1,r_2](z)$ together with their operator product expansions (OPEs)
\begin{align}
    \label{JJopeprim}
    &J^a[r_1,r_2](z)J^b[s_1,s_2](w)\sim \frac{if^{abc}}{z-w}J^c[r_1+s_1,r_2+s_2](w)\,
    \\
    \label{JJtopeprim}
    &J^a[r_1,r_2](z)\tilde{J}^b[s_1,s_2](w)\sim \frac{if^{abc}}{z-w}\tilde{J}^c[r_1+s_1,r_2+s_2](w)
    \, ,
\end{align}
where we use the standard notation $\sim$ to denote the restriction to the singular terms.
These generators are Lorentz boost eigenstates, 
and they can be thought of as conformal primaries of a two-dimensional CFT. 
From eqs.~\eqref{JJopeprim} and \eqref{JJtopeprim}, the momentum eigenstates are seen to have the OPEs
\begin{align}
    \label{JJope}
    &J^a(\tilde\lambda_1,z_1)J^b(\tilde\lambda_2,z_2)\sim \frac{if^{abc}}{\spa1.2} J^c(\tilde\lambda_1+\tilde\lambda_2,z_2)
    \\
    \label{JJtope}
    &J^a(\tilde\lambda_1,z_1)\tilde{J}^b(\tilde\lambda_2,z_2)\sim \frac{if^{abc}}{\spa1.2} \tilde{J}^c(\tilde\lambda_1+\tilde\lambda_2,z_2)
    \,,
\end{align}
where we used $\spa i.j := \epsilon_{\alpha\beta}\lambda_i^\alpha\lambda_j^\beta=z_i-z_j$.
These OPEs capture the holomorphic collinear behavior of the self-dual theory and, for  computational purposes, they are most conveniently
organized in terms of $J[k]$. They are given by, using eqs.~\eqref{JJopeprim} and \eqref{JJtopeprim},
\begin{align}
    \label{JJmodeope}
    &J^a[k_1](z_1)J^b[k_2](z_2)\sim \frac{if^{abc}}{\spa1.2}
    \frac{1}{k_1!}\left(\tilde\lambda_1^{\dot\alpha}\frac{\partial}{\partial\tilde\lambda_2^{\dot\alpha}}\right)^{k_1}
    J^c[k_1+k_2](z_2)
    \\
    \label{JJtmodeope}
    &J^a[k_1](z_1)\tilde{J}^b[k_2](z_2)\sim \frac{if^{abc}}{\spa1.2} 
    \frac{1}{k_1!}\left(\tilde\lambda_1^{\dot\alpha}\frac{\partial}{\partial\tilde\lambda_2^{\dot\alpha}}\right)^{k_1}
    \tilde{J}^c[k_1+k_2](z_2)
    \,.
\end{align}
The derivatives lower the powers of $\tilde\lambda_2$ by $k_1$ powers while introducing $k_1$ powers of $\tilde\lambda_1$, making the right-hand sides of eqs.~\eqref{JJmodeope}--\eqref{JJtmodeope} have the same $k_i$ powers of the components of $\tilde\lambda_i$ as the left-hand sides.

Correlation functions of CA operators can be computed using OPE techniques familiar from two-dimensional CFT, once a conformal block has been chosen.\footnote{\linespread{0.9}\selectfont Following Ref.~\cite{Costello:2022wso}, a conformal block specifies a consistent set of correlation functions of CA operators compatible with their OPEs.} 
In the construction of Ref.~\cite{Costello:2022wso}, these conformal blocks are in one-to-one correspondence with local gauge-invariant operators of the four-dimensional theory. Correlators evaluated in the block associated with a local operator $\mathcal{O}$ yield the four-dimensional form factors of $\mathcal{O}$.

For the gauge theory, translation covariance allows us to write such a form factor as
\begin{equation}
\langle C_1(\tilde\lambda_1,z_1)\cdots C_n(\tilde\lambda_n,z_n)
|O(x)\rangle
=
e^{iP\cdot x}
\langle C_1(\tilde\lambda_1,z_1)\cdots C_n(\tilde\lambda_n,z_n)
|O(0)\rangle
\,,
\end{equation}
where the $C_i(\tilde\lambda_i,z_i)$ collectively denote the momentum eigenstates $J$ and ${\tilde J}$, and $P$ is the total momentum of the scattering states. We will work with $O(0)$, for simplicity, and drop the explicit dependence on $x=0$ from our notation. The momentum eigenstates can then be expanded into energy modes as in Eq.~\eqref{eq:Jexpansion} and its analog for ${\tilde J}$, giving
\begin{equation}
\langle C_1(\tilde\lambda_1,z_1)\cdots C_n(\tilde\lambda_n,z_n)
|O\rangle
=
\sum_{\substack{k_1,\dotsc,k_n\geq0 \\ \Delta_1+\cdots+\Delta_n=-\Delta_O}}
\langle C_1[k_1](z_1)\cdots C_n[k_n](z_n)
|O\rangle
\ .
\end{equation}
Each term in the sum on the right-hand side is a correlation function of the chiral algebra operators in the conformal block defined by $O$. 
The $\Delta_i$ are the spacetime scaling dimensions of $C_i[k_i]$, and they depend on the value of $k_i$. 
The constraint equation $\Delta_1+\cdots+\Delta_n=-\Delta_O$ is a consequence of scale invariance. For correlation functions of composite operators
in CFTs scale invariance correlates the position dependence with the scaling dimension of the operators; here however, the operators are independent of spacetime position so this reduces to conservation of scaling dimensions.
Since~\cite{Costello:2022wso}
\begin{equation}
\Delta\bigl(J^a[k]\bigr)=-k,
\qquad
\Delta\bigl(\widetilde J^a[k]\bigr)=-2-k \ ,
\end{equation}
the conservation of scaling dimensions in turn relates the values of the $k_i$ with the scaling dimension $\Delta_O$ of the ${\cal O}$. 

For a scalar operator $O(0)$, Lorentz invariance 
requires that all spinors $\lambda$ and $\tilde\lambda$ appear in the form factor only through the invariant pairings 
\begin{equation}
\spa i.j=z_i-z_j 
\quad
\text{and}
\quad
\spb i.j:=-\epsilon_{\dot\alpha\dot\beta}\tilde\lambda_i^{\dot\alpha}\tilde\lambda_j^{\dot\beta} \ ,
\end{equation}
respectively. 
The expansion \eqref{eq:modeex} of modes $C[k]$ into generators $C[t_1,t_2]$ then implies that any correlation function of $C[k_i]$'s will have $k_i$ powers of the components of $\tilde\lambda_i$ appearing in numerators only. 
So, all CA correlation functions will contain nonnegative powers of square brackets $\spb i.j$~\footnote{\linespread{0.9}\selectfont This also implies a selection rule that the sum of mode numbers must be even.}, and poles can only occur in the angle brackets $\spa i.j$. 
The OPEs of the chiral algebra constrain the singular behavior of its correlation functions by determining the residues of the poles $\spa i.j$, with the residues being lower-point correlators. This is akin to the usual factorization properties of scattering amplitudes.

Here, however, the asymptotic behavior of correlation functions at $z_i=\infty$ is also constrained.  Indeed, the modes $C_i[k_i](z_i)$ have a holomorphic two-dimensional conformal weight determined by $k_i$,
\begin{equation}
h\bigl(J^a[k]\bigr)=1-\frac{k}{2},
\qquad
h\bigl(\widetilde J^a[k]\bigr)=-1-\frac{k}{2} \ .
\end{equation}
These weights in turn dictate the asymptotic behavior of correlation functions when $z_i\to\infty$, 
\begin{equation}
\langle C_1[k_1](z_1)\cdots C_n[k_n](z_n)|O\rangle \big|_{z_i\rightarrow\infty} \sim {\cal O}(z_i^{-2h_i}) \ ,
\end{equation}
where $h_i\equiv h(C_i[k_i]z_i)$.
When the chosen insertion has positive holomorphic weight, the correlator vanishes at infinity. Its finite-pole singularities, determined by the OPEs and lower-point correlators, then fix it uniquely: two expressions with the same singular parts differ by a polynomial, which must vanish if it vanishes at infinity.
This falloff data is then  sufficient to explicitly reconstruct correlators from their OPE singularities and to prove their uniqueness given a seed correlator, thus bootstrapping the $n$-point correlators from the lower-point ones~\cite{Costello:2022wso} and making this construction more 
constraining than an approach based on factorization alone.

For special anomaly-free twistorial gauge theories, symmetry constraints and associativity have been used to determine the extended OPEs at arbitrary loop order~\cite{Fernandez:2024qnu}. Anticipating seed two-loop four-point all-plus amplitudes derived in Sec.~\ref{sec:FourPointTwoloopAmpitude}, we 
now proceed to use these ideas to derive the $n$-point two-loop all-plus amplitudes in GR$_\Psi$ theory.

\subsection{The $F^3$ form factor and its amplitude limit}
\label{subsec:F3_form_factor}

We will use the KLT relations to obtain the two-loop all-plus amplitudes in GR$_\Psi$ theory from those of YM-matter theory deformed by an $F^3$ operator.
Since the KLT relations use color-ordered amplitudes, it is convenient to work in a trace basis.
Thus, we write $F_{\mu\nu}=F_{\mu\nu}^a t^a$, where $t^a$ are the fundamental generators of $SU(N_c)$, $N_c$ is the number of colors, and $\tr$ denotes a color trace. The notation $F^3$ refers to the cubic gauge-invariant operator $\tr(F_\mu{}^\nu F_\nu{}^\rho F_\rho{}^\mu)$, with its overall normalization fixed below. The operators $F_\pm^3$ use the same contraction with $F$ replaced by $F_\pm$, where \begin{equation} \label{eq:F3_projections} (F_\pm)_{\mu\nu} = \frac{1}{2}(F_{\mu\nu}\pm\tilde F_{\mu\nu})\,. \end{equation} The dual is defined using the four-dimensional Levi--Civita tensor $\epsilon_{\mu\nu\rho\sigma}$ as
\begin{equation}
\tilde F_{\mu\nu} = \frac{i}{2}\epsilon_{\mu\nu\rho\sigma}F^{\rho\sigma}
\end{equation}

Separating $F$ into self-dual and anti-self-dual components $F_+$ and $F_-$, respectively, as $F=F_++F_-$, the all-plus tree-level form factor of a single $F_-^3$ insertion vanishes in the self-dual theory. In other words,
\begin{equation}
A_n^{F^3}(1^+,2^+,\dotsc,n^+) = A_n^{F_+^3}(1^+,2^+,\dotsc,n^+)
\,.
\end{equation}
This equality is most apparent in the self-dual theory, as defined by Chalmers and Siegel \cite{Chalmers:1996rq} and reviewed in Sec.~\ref{subsec:review_chiral_algebra}, where minus-helicity gluons are associated with an anti-self-dual two-form $B$ and positive-helicity gluons are from the gauge field $A$. The field $B$ plays the role of $F_-$, so the nonzero tree amplitudes with a single $F_-^3$ insertion have three negative-helicity gluons in this self-dual theory. A single $F_+^3$ insertion produces only all-plus tree amplitudes in the same theory.\footnote{\linespread{0.9}\selectfont This convention that $F_+$ sources positive-helicity gluons and $B$ sources the negative-helicity ones is opposite to refs.~\cite{Dixon:2004za,Broedel:2012rc}, originating from different choices for the Levi-Civita tensor.}

Here $A_n^{F^3}$ denotes the tree-level color-ordered form factor for the cyclic ordering $(1,\ldots,n)$, with overall coupling factors suppressed. Its phase and normalization are fixed by the three-point seed \begin{equation} \label{eq:F3_normalization} A_3^{F^3}(1^+,2^+,3^+) = i\,\frac{s_{12}s_{23}s_{31}}{\spa1.2\spa2.3\spa3.1} = -i\,\spb1.2\spb2.3\spb3.1\,. \end{equation} The color-ordered correlators in Appendix~\ref{app:F3_derivation} are written without this overall factor of $i$. The all-plus $n$-gluon form factor, derived there using the CAB, is
\begin{equation}
\label{eq:F3}
A_n^{F^3}(1^+,2^+,\dotsc,n^+)
=i\sum_{k=3}^6\sum_{1\leq i_1<\cdots<i_k\leq n} \frac{T_k(i_1,\dotsc,i_k)}{\spa1.2\spa2.3\cdots\spa n.1}
\,,
\end{equation}
where the functions $T_k$ are
\begin{align}
    \label{eq:T3}
    &T_3(i,j,k) = s_{ij}s_{jk}s_{ki}
    \\
    \label{eq:T4}
    &T_4(i,j,k,l) = s_{ij}s_{jk}s_{li} + s_{ij}\spa j.k\spb k.i\spa i.l\spb l.j
    + \mathcal{C}(i,j,k,l)
    \\
    \label{eq:T5}
    \begin{split}
    &T_5(i_1,\dotsc,i_5) = s(i_1,i_2)s(i_3,i_4)s(i_5,i_1)
            + s(i_1,i_2)\spa{i_5}.{i_1}\spb{i_1}.{i_4}\spa{i_4}.{i_3}\spb{i_3}.{i_5}
            \\
            &\hspace{2.5cm}+s(i_5,i_1)\spa{i_1}.{i_2}\spb{i_2}.{i_4}\spa{i_4}.{i_3}\spb{i_3}.{i_1}
            + \spa{i_5}.{i_1}\spb{i_1}.{i_4}\spa{i_4}.{i_3}\spb{i_3}.{i_1}\spa{i_1}.{i_2}\spb{i_2}.{i_5}
            \\
            &\hspace{2.5cm}+ \mathcal{C}(i_1,i_2,i_3,i_4,i_5)
    \end{split}
    \\
    \begin{split}
    &T_6(i_1,\dotsc,i_6) = 
    -\spa{i_1}.{i_2}\spa{i_3}.{i_4}\spa{i_5}.{i_6}
    \big(\spb{i_1}.{i_2}\spb{i_3}.{i_4}\spb{i_5}.{i_6} 
        + \spb{i_6}.{i_2}\spb{i_3}.{i_4}\spb{i_5}.{i_1}
        + \spb{i_1}.{i_3}\spb{i_2}.{i_4}\spb{i_5}.{i_6}
        \\
        &\hspace{2.5cm}
        + \spb{i_1}.{i_2}\spb{i_3}.{i_5}\spb{i_4}.{i_6}
        + \spb{i_6}.{i_3}\spb{i_2}.{i_4}\spb{i_5}.{i_1} 
        + \spb{i_6}.{i_2}\spb{i_3}.{i_5}\spb{i_4}.{i_1}
        \\
        &\hspace{2.5cm}
        + \spb{i_1}.{i_3}\spb{i_2}.{i_5}\spb{i_4}.{i_6} 
        + \spb{i_6}.{i_3}\spb{i_2}.{i_5}\spb{i_4}.{i_1}
        \big)
        + (i_2\,i_3\,i_4\,i_5\,i_6\, i_1)
        \,.  
    \end{split}
      \label{eq:T6} 
\end{align}
In Eqs.~\eqref{eq:T4} and \eqref{eq:T5}, $+\mathcal{C}(j_1,\ldots,j_m)$ instructs one to add the images of the entire preceding expression under all nonidentity cyclic permutations of the listed arguments.
In contrast, $+ (i_2\,i_3\,i_4\,i_5\,i_6\, i_1)$ in Eq.~\eqref{eq:T6} instructs one to add only the image under the single cyclic shift $i_1\to i_2\to\cdots\to i_6\to i_1$.

The above expression is for the form factor of $F^3$, whose operator insertion supplies the total external momentum $P=\sum_{i=1}^n p_i$, which need not vanish. Setting $P=0$ gives the momentum-conserving all-plus tree amplitude with a single $F^3$ insertion~\cite{Broedel:2012rc}, which we use in the KLT construction below.

At four points, only the $T_3$ and $T_4$ terms contribute to Eq.~\eqref{eq:F3}. For nonzero operator momentum $P=p_1+p_2+p_3+p_4$, the form factor is
\begin{equation}
A_4^{F^3}(1^+,2^+,3^+,4^+;P)
=i\,\frac{N_4}{\spa1.2\spa2.3\spa3.4\spa4.1}\,,
\label{eq:F3_fourpoint_formfactor}
\end{equation}
where
\begin{equation}
\begin{split}
N_4={}&s_{12}s_{23}s_{31}+s_{12}s_{24}s_{41}
+s_{13}s_{34}s_{41}+s_{23}s_{34}s_{42}
\\
&+\sum_{\mathrm{cyc}(1234)}
\left(s_{12}s_{23}s_{41}
+s_{12}\spa2.3\spb3.1\spa1.4\spb4.2\right)\,.
\end{split}
\label{eq:F3_fourpoint_numerator}
\end{equation}
The cyclic sum runs over the four orderings $(1234)$, $(2341)$, $(3412)$, and $(4123)$. The first line collects the four $T_3$ contributions, while the second line is $T_4(1,2,3,4)$. For $P\neq0$, the usual momentum-conserving four-point identities among the Mandelstam invariants do not apply. In the amplitude limit $P=0$, however, $s=s_{12}=s_{34}$, $t=s_{23}=s_{14}$, and $u=s_{13}=s_{24}$ satisfy $s+t+u=0$, and the numerator reduces to $N_4=-2stu$. Thus
\begin{equation}
\left.A_4^{F^3}(1^+,2^+,3^+,4^+;P)\right|_{P=0}
=-2i\,\frac{stu}{\spa1.2\spa2.3\spa3.4\spa4.1}\,,
\label{eq:F3_fourpoint_amplitude}
\end{equation}
where the phase and normalization are fixed by Eq.~\eqref{eq:F3_normalization}, with overall coupling factors suppressed.

\subsection{The $n$-point amplitude from KLT}

Unlike in SDYM, all form factors of the operator deforming SDGR towards GR, as well as of other operators deforming the action, are not gauge invariant, as that deformation is not invariant under diffeomorphisms. 
Its integral over the entire spacetime is, however, invariant. We therefore compute momentum-conserving amplitudes with insertions of the resulting vertices, 
rather than form factors. The same strategy can be applied to gauge theories, where it leads to amplitudes in a theory deformed by gauge-invariant local operators.
These amplitudes are integrals of the form factors of these operators. We will use the KLT relations to obtain the amplitudes of $R^3$-deformed GR from the amplitudes of $\tr[F^3]$-deformed YM theory. 

In this subsection, we use the form of the KLT kernel given in Ref.~\cite{Bern:1998sv}:
\begin{equation}
\begin{aligned} M_n(1,2,\dotsc,n) ={}& i\,(-1)^{n+1}\sum_{\sigma\in S(2,\dotsc,n-2)}
A_n(1,\sigma,n-1,n)
\\ &\times\sum_{(\tau,\rho)\in\mathcal{P}}\sigma\!\left[f(\tau)\bar f(\rho)\right]
A_n(\sigma\circ\tau,1,n-1,\sigma\circ\rho,n)
\,,
\end{aligned}\label{eqn:Vertex}
\end{equation}
where
\begin{align}
\mathcal{P} = 
\begin{cases}
S(2,\dotsc,n/2)\times S(n/2+1,\dotsc,n-2),
& \text{for even}~n\,,
\\
S(2,\dotsc,(n{-}1)/2)\times S((n{+}1)/2,\dotsc,n-2),
& \text{for odd}~n\,,
\end{cases}
\end{align}
and $S(X)$ is the symmetric group on the set $X$. The functions $f$ and $\bar{f}$ are given by
\begin{align}
\begin{split}
f(i_1,\dotsc,i_j) &= 
s(1,i_j)\prod_{m=1}^{j-1}\left(s(1,i_m)+\sum_{k=m+1}^jg(i_m,i_k)\right)
\\
\bar{f}(i_1,\dotsc,i_l) &= s(i_1,n{-}1)\prod_{m=2}^{l}
\left(s(i_m,n{-}1)+\sum_{k=1}^{m-1}g(i_k,i_m)\right)
\,,
\end{split}
\end{align}
with $f(\emptyset)={\bar f}(\emptyset)=1$, i.e. for $j, l=0$, and
\begin{align}
g(i,j) =\label{eq:KLT_g}
\begin{cases}
s(i,j)\equiv s_{ij}, & i>j
\\
0, & \text{otherwise}\,.
\end{cases}
\end{align}
Following Ref.~\cite{Bern:1998sv}, the functions $f(\tau)$ and $\bar f(\rho)$ are first evaluated using the unpermuted labels, including the ordering condition $i>j$ in Eq.~\eqref{eq:KLT_g}. The outer permutation $\sigma$ then relabels every Mandelstam invariant in the resulting polynomial according to 
$ \sigma[s_{ij}]=s_{\sigma(i)\sigma(j)}$,  with $\sigma$ fixing $1,n-1,n$. 

The KLT double copy of two gauge-theory amplitudes, each with a single $F^3$ insertion, contains dilaton and axion exchange in addition to the pure-graviton contribution~\cite{Broedel:2012rc}. With matched three-point normalizations, the four- and five-point identical-helicity amplitudes with a single $R^3$ insertion are $5/2$ times the KLT result. In these amplitudes, axion exchange cancels, while the contribution from two $\phi R^2$ insertions, where $\phi$ is the dilaton, is $-3/2$ times the KLT result, accounting for the difference. The persistence of all-plus proportionality through seven points was explicitly checked in Ref.~\cite{Dunbar:2017dgp}. We fix the overall normalization by matching to the four-point $R^3$ seed.  We note that this simple relation between amplitudes with purely gravitational exchanges and those that include dilaton exchanges makes it straightforward to use the double copy, given that the double copy naturally includes the dilaton.

To assemble the $n$-point two-loop all-plus gravity amplitude, we first insert the functions $T_3,\ldots,T_6$ in Eqs.~\eqref{eq:T3}--\eqref{eq:T6} into the gauge-theory expression~\eqref{eq:F3} and set the operator momentum $P=\sum_{i=1}^n p_i$ to zero. We then substitute the resulting momentum-conserving amplitudes $A_n^{F^3}$ for both gauge-theory amplitudes $A_n$ in the KLT relation~\eqref{eqn:Vertex}, relabeling the external legs according to the ordering of each factor and using the permutation sums and kernel functions defined above. 
The IR-subtracted, UV-renormalized amplitude in the chiral theory is therefore
\begin{equation}
\Mcal^{(2),\rm ren}_{n,\rm chiral}(1^+,\ldots,n^+)
= \frac{5}{2}\,\alpha\left(\frac{\kappa}{2}\right)^{n-2}
\left. M_n(1,\ldots,n)\right|_{A_n\to A_n^{F^3},\,P=0}
\,,\qquad n\geq4\,,
\label{eq:chiral_allplus_KLT}
\end{equation}
where $M_n$ is the KLT expression in Eq.~\eqref{eqn:Vertex}, with both gauge-theory factors replaced by the all-plus $F^3$ amplitudes as described above.
The coefficient $\alpha$ is given explicitly in Eq.~\eqref{eq:chiral_alpha}. 
The factor $5/2$ converts the double-copy result, which includes dilaton exchange, to the normalization of the pure-gravity amplitude with a single $R^3$ insertion, while $\alpha$ is fixed by the explicit four-point calculation in \sect{sec:FourPointTwoloopAmpitude}. The factor $(\kappa/2)^{n-2}$ restores the conventional gravitational coupling of an $n$-point tree amplitude; together with $\alpha\propto(\kappa/2)^4$, it gives the two-loop scaling $(\kappa/2)^{n+2}$. 

To confirm that the bootstrap and insertion into the KLT formula are done correctly, we explicitly checked the leading soft limits and crossing through 13 points\footnote{\linespread{0.9}\selectfont To reach 13 points we used Codex to write a C++ parallelized version with appropriate caching submitted to a supercomputer.}.  Our crossing check consisted of sequentially checking Bose symmetry for pairs of swaps equivalent to checking all pairs of swaps.

At four points, inserting the gauge-theory amplitude~\eqref{eq:F3_fourpoint_amplitude} and its relabeled ordering into the KLT relation gives the double-copy amplitude
\begin{align}
M_4^{\mathrm{DC}}(1^+,2^+,3^+,4^+)
&=-i\,s_{12}\,A_4^{F^3}(1,2,3,4)\,A_4^{F^3}(2,1,3,4)
\nonumber\\
&=4i\Tcal^2stu\,.
\label{eq:F3_fourpoint_KLT}
\end{align}
All external helicities in the two gauge-theory factors are positive. Applying the $5/2$ conversion factor and restoring $\alpha(\kappa/2)^2$, the chiral-theory amplitude is then
\begin{align}
\Mcal^{(2),\rm ren}_{4,\rm chiral}(1^+,2^+,3^+,4^+)
&=\frac{5}{2}\,\alpha\left(\frac{\kappa}{2}\right)^2
M_4^{\mathrm{DC}}(1^+,2^+,3^+,4^+)
\nonumber\\
&=10i\alpha\left(\frac{\kappa}{2}\right)^2\Tcal^2stu\,.
\label{eq:chiral_KLT_fourpoint_match}
\end{align}
Thus, the numerical factor $10=(5/2)\times4$ agrees with Eq.~\eqref{eq:AmplChiral} in \sect{sec:FourPointTwoloopAmpitude}. With this definition, $10\alpha(\kappa/2)^2$ is precisely the coefficient of $i\Tcal^2stu$ in the renormalized four-point amplitude of Sec.~\ref{sec:FourPointTwoloopAmpitude} after setting $N_{\!f}=2$; the logarithmic terms vanish for this chiral matter content. 

There is a special value of the overall factor $\alpha = 0$, which would result in an especially simple amplitude, namely,
\begin{equation}
\Mcal^{(2),\rm ren}_{n,\rm chiral}(1^+,\ldots,n^+) = 0 \, .
\end{equation}
As noted in \sect{sec:FourPointTwoloopAmpitude}, the coefficient $\alpha$ does not run, so setting it to any value, including 0, leaves it unchanged.  
The recursion relation of Ref.~\cite{Dunbar:2017dgp} establishes a one-to-one correspondence between the three- and four-point all-plus amplitudes at two loops. Setting the four-point amplitude to zero, therefore, also eliminates the three-point amplitude.

One might wonder whether there is a reason for the choice $\alpha = 0$ as natural. Supersymmetry, for example, would enforce this choice~\cite{Grisaru:1976vm, Grisaru:1977px, Parke:1985pn, Kunszt:1985mg}. Refs.~\cite{Mago:2021wje, Ren:2022sws} showed that the $(+++)$ vertex sourced by $R^3$ introduces an anomaly to the associativity of a tree-level current algebra in the CCFT framework (distinct from the associativity of the CA \cite{Costello:2022wso,Bittleston:2022jeq}) coupled to full GR. This might suggest that preserving associativity of this current algebra may be a reason to set $\alpha=0$. However, the analyses of refs.~\cite{Mago:2021wje, Ren:2022sws} did not incorporate loop-level OPEs into their associativity computations, so it is not clear whether $R^3$ obstructs associativity when considered as a source for two-loop amplitudes. From explicit amplitudes, we explicitly checked that associativity of the collinear limits holds in the all-plus configuration. 
This is perhaps not unexpected, since this factorization limit is proportional to the tree-level single-minus amplitude, which vanishes identically.

\section{Four-point two-loop all-plus helicity amplitude}
\label{sec:FourPointTwoloopAmpitude}

In this section, we present the one- and two-loop all-plus four-graviton amplitudes for gravity minimally coupled to a chiral fermion. 
At two loops, the IR-subtracted, UV-renormalized four-graviton amplitude in Einstein's gravity minimally coupled to matter is assembled as 
\begin{align}
\label{eq:2-loop_remainder_pppp}
\Mcal^{(2),\rm ren}_4 = \Mcal^{(2)}_4 \!
- \Gcal_4 \,  \Mcal^{(1)}_4 
+ \Mcal^{(1)}_{4,{\rm GB}} 
+ \Mcal^{(0)}_{4,{\rm GB^2}} 
+ \Mcal^{(0)}_{4,{\rm R}^3}\, ,
\end{align}
where the terms on the right-hand side are the bare two-loop amplitude $\Mcal^{(2)}_4$, its infrared subtraction $\Gcal_4 \,  \Mcal^{(1)}_4$ (using the vanishing of the tree-level all-plus amplitude), the one-loop single Gauss--Bonnet insertion, the tree-level double Gauss--Bonnet insertion, and the $R^3$ counterterm insertion. 
We present these contributions and the renormalized result in turn. 
The factorization of the IR singularities requires us to use one-loop amplitudes, including their dependence on the regulator; they remove IR artifacts from the two-loop amplitude. 
We leave the number of fermion states as a free parameter to expose the cancellation of the logarithms, which results in a purely rational subtracted and renormalized amplitude for Einstein's gravity coupled to a single chiral fermion. 

The calculations here use the standard tools including generalized unitarity methods~\cite{Bern:1994zx, Bern:1994cg, Bern:1997sc, Britto:2004nc, Bern:2004cz, Bern:2007ct}, color-kinematics duality, and the double-copy construction~\cite{Kawai:1985xq, Bern:2008qj, Bern:2010ue, Bern:2019prr}, together with modern multi-loop integration techniques~\cite{Chetyrkin:1981qh, Tkachov:1981wb, Kotikov:1990kg, Bern:1993kr, Remiddi:1997ny, Gehrmann:1999as, Henn:2013pwa}. Using these tools, we have computed the full two-loop four-graviton amplitude for the theory defined by Eq.~(\ref{LagrangianChiral}), with generic external polarization tensors.  Here, we focus only on the all-plus helicity amplitude, relevant for the CA. The same setup was used to compute all two-loop four-graviton helicity amplitudes; those results will be reported elsewhere~\cite{FuturePaper}.

\begin{figure}[tb]
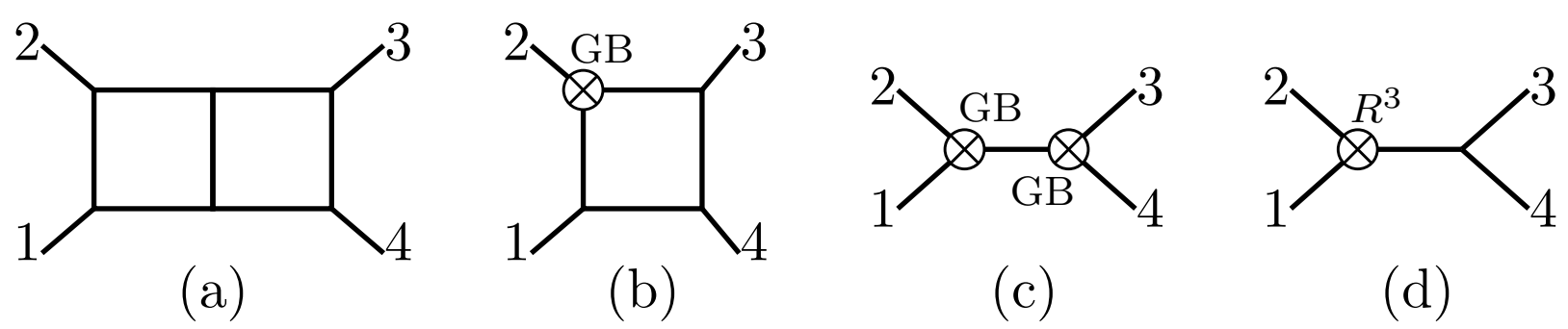

\centering
\representativeContributionsTikZ
\vskip -.2 cm 
\caption{Representative diagrams of 
(a) the two-loop bare,  
(b) the one-loop single Gauss--Bonnet insertion, 
(c) the tree-level double Gauss--Bonnet counterterm insertion and 
(d) the $R^3$ counterterm insertion. }
\label{fig:TwoloopCounterterm}
\end{figure}

\subsection{The bare two-loop four-point all-plus helicity amplitudes in \GRmatterPlus}
\label{subsec:BareTwoloopAmpitude}

We construct the two-loop integrand via the generalized unitarity method~\cite{Bern:1994zx, Bern:1994cg, Bern:1997sc, Bern:2004cz, Bern:2007ct}. The gravitational tree amplitudes are obtained as double copies~\cite{Bern:2010ue, Bern:2019prr} of suitable YM amplitudes~\cite{Bern:2008qj,Edison:2020ehu}.\footnote{\linespread{0.9}\selectfont The graviton polarization tensors are naturally expressed in terms of products of transverse (i.e. $\varepsilon_i \cdot p_i=0$) and null (i.e. $\varepsilon_i^2 =0$) Yang-Mills polarization vectors $\varepsilon_i^{\mu_i\nu_i} = \varepsilon_i^{\mu_i}\varepsilon_i^{\nu_i}$.}
The tree amplitudes with four or more fermions produced by the double copy include dilaton and antisymmetric tensor exchanges. 
Rather than introduce additional ghost fields~\cite{Johansson:2014zca} to isolate the graviton exchange, we exploit the simplicity of the four-fermion tree amplitude and derive it directly using the Feynman rules from the Lagrangian~\eqref{LagrangianChiral}.
%
We merge the resulting expressions for the generalized cuts using the construction in Ref.~\cite{Bern:2024vqs}.

\subsubsection{Integrand construction}

\begin{figure}[tb]
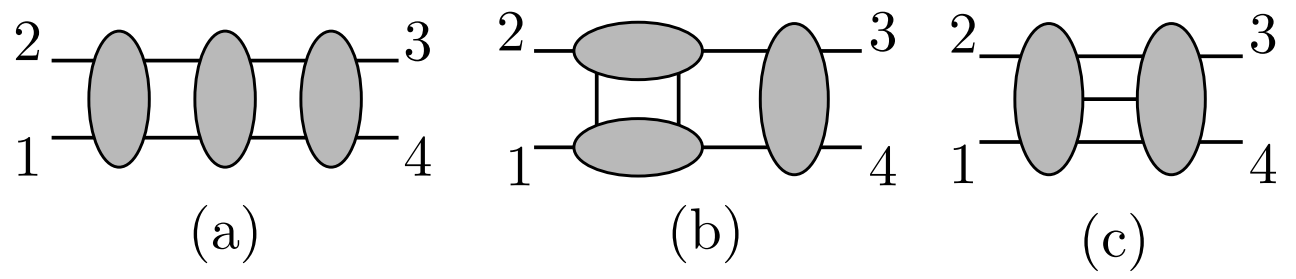

\centering
\spanningCutsTikZ
\vskip -.2 cm 
\caption{\label{span2L}
The three independent spanning cuts, up to relabelings of external legs. The exposed lines are on shell.}
\end{figure}

Up to relabelings of external legs and field assignments, the usual two-particle cut determines the four-point one-loop amplitude in \GRmatterPlus{} theory. The three independent spanning cuts determining the four-point two-loop amplitudes are shown in Fig.~\ref{span2L}.

With tree amplitudes organized~\cite{Kosmopoulos:2020pcd} to obey the generalized Ward identities, the evaluation of unitarity cuts is straightforward.
We use $D_s\equiv\eta^\mu{}_\mu$ to count the number of spacetime dimensions the graviton lives in, so that the number of graviton physical states is $D_s (D_s - 3)/2$.  Generally, it is useful to treat $D_s$ as distinct from the number of spacetime dimensions $D$; here, however, we present our results in the 't Hooft-Veltman (HV) scheme with $D=D_s=4-2\eps$. In the explicit two-loop computation, it is useful to track the fermion contributions relative to the graviton, which we do by tracking the number of Dirac traces of the fermions.
To further simplify the integrand, we choose the cyclic gauge 
$
\varepsilon_1 \cdot p_{2}{=}
\varepsilon_2 \cdot p_{3}{=}
\varepsilon_3 \cdot p_{4}{=}
\varepsilon_4 \cdot p_{1}{=}0.
$

\subsubsection{Tensor projection}

The resulting integrand contains contractions between external polarization tensors and loop momenta, which make integration-by-parts (IBP) reduction difficult by introducing unnecessary irreducible numerators. Tensor projection~\cite{Garland:2002ak, Peraro:2020sfm} addresses this challenge by systematically trading such tensor integrals for scalar integrals (integrals expressible only in terms of momentum invariants). 
%
In cyclic gauge, 
there exist 56 tensor structures, $T_a$, that have the appropriate homogeneity in the graviton polarizations, $T_a=
\varepsilon_1^{\mu_1\nu_1}
\varepsilon_2^{\mu_2\nu_2}
\varepsilon_3^{\mu_3\nu_3}
\varepsilon_4^{\mu_4\nu_4}(T_a)_{\mu_1\nu_1,\mu_2\nu_2,\mu_3\nu_3,\mu_4\nu_4}, a=1,\ldots,56$ and depend only on external momenta.

A priori, the tensors $T_a$ do not form an orthogonal basis and give rise to a nontrivial Gram matrix $G_{ab}= T_a \cdot T_b\equiv \sum_{{\rm states}} \overline{T}_a\, T_b$. Here $\overline{T}_a$ is obtained from $T_a$ by replacing each external polarization $\varepsilon_i$ with its dual $\overline{\varepsilon}_i$; for real momenta, these dual polarizations may be chosen as complex conjugates. The inner product is defined by summing over a complete set of physical states for each external graviton (in cyclic gauge),
\begin{align}
    \sum_{{\rm states}} \overline{\varepsilon}^{\mu \nu }_i  \varepsilon^{\rho \sigma }_i = \frac{1}{2}
    \Bigg(& 
    \Pi^{\mu \rho}_{p_i,p_{i+1}}\Pi^{\nu \sigma}_{p_i,p_{i+1}}
    +
    \Pi^{\mu \sigma}_{p_i,p_{i+1}}\Pi^{\nu \rho}_{p_i,p_{i+1}}
    -\frac{2}{D_s-2} \Pi^{\mu \nu}_{p_i,p_{i+1}}\Pi^{\rho \sigma}_{p_i,p_{i+1}}
    \Bigg)\,,
\end{align}
expressed in terms of the Yang-Mills light-cone projectors $\Pi^{\mu \nu}_{k,q} {=} \, \eta^{\mu \nu} {-}\, \frac{k^\mu q^\nu + k^\nu q^\mu}{k\cdot q}$ where we identify $p_5\,{=}\,p_1$. 

Let $V$ denote the space spanned by the 56 basis tensors $T_a$. With this inner product, the resolution of the identity acts on any $X\in V$ as
\begin{equation}
\mathbb{I}_V X = \sum_{a,b} T_a\, (G^{-1})_{ab}\, (T_b\cdot X) = X\,.
\end{equation}
Then, the tensor-projected amplitude is 
\begin{equation}
\mathcal{M} = \sum_a c_a\, T_a \ ,\quad \text{with}\quad
c_a= \sum_b (G^{-1})_{ab}\, (T_b\cdot\mathcal{M})\,.
\end{equation}
The inner products $T_a\cdot\mathcal{M}$ are evaluated via graviton state sums, as described above, yielding an integrand where the complete polarization dependence has been factorized at the price of introducing nontrivial denominators through the matrix elements of the inverse of the inner products of tensor basis elements.

\subsubsection{IBP Reduction and integral evaluation}

We organize the resulting tensor-projected integrand in an inverse-propagator basis and find three integral families (modulo relabeling of external legs). In this
form, it is ideally suited for IBP reduction~\cite{Chetyrkin:1981qh, Tkachov:1981wb, Laporta:2000dsw}. Modern automated computer programs~\cite{Lange:2025fba, Guan:2024byi, FIRE7, Maierhofer:2017gsa, Smirnov:2023yhb, Smirnov:2008iw} make this straightforward despite the high power counting of gravitational interactions. Using integral symmetries and identifying equivalent integrals across families, we obtain the 30 global master integrals of Ref.~\cite{Abreu:2020lyk}. 
These master integrals were evaluated analytically in earlier work~\cite{Smirnov:1999gc, Tausk:1999vh, Smirnov:1999wz, Anastasiou:2000kp}. Here we recomputed their expansions in the dimensional regulator parameter $\epsilon=(4-D)/2$ using differential equations~\cite{Kotikov:1990kg, Bern:1993kr, Remiddi:1997ny, Gehrmann:1999as} in canonical form~\cite{Henn:2013pwa, Henn:2013woa}, expressing the results in terms of iterated integrals.

Thus, we obtain the one-loop four-graviton all-plus helicity amplitudes in \GRmatterPlus{} for a general number $N_{\!f}$ of fermionic degrees of freedom,
\begin{align}
\hspace{-.5cm}
\Mcal^{(1)}_{4}(1^+, 2^+,3^+,4^+) =  i\, \left(\frac{\kappa}{2}\right)^4  L^{(1)} (\epsilon) \,
\Tcal^2  
\left[
\frac{ N_{\!f} -2 }{240}
(s^2{+}t^2{+}u^2) 
+ \Ocal(\epsilon)
\right]\!.
\hspace{-.5cm}
\label{FourPointOneLoopAmpl} 
\end{align}
Here, $L^{(1)}(\epsilon)$ is the one-loop normalization factor \eqref{LNormalization}. This amplitude is both infrared and ultraviolet finite, because the all-plus tree-level amplitude vanishes, and the potential one-loop counterterm is evanescent~\cite{tHooft:1974toh}, respectively. 
The $\Ocal(\eps)$ terms are important for the infrared divergence subtraction at two loops and are given in the ancillary file. 
Our $N_{\! f}=0$ results agree with those in the ancillary file of Ref.~\cite{Abreu:2020lyk} and with the $\Ocal(\epsilon^0)$ terms of Ref.~\cite{Bern:1998sv}. 

The bare two-loop all-plus helicity four-graviton amplitude in \GRmatterPlus{} for general $N_{\!f}$ is
\begin{align}
\label{eq:2-loop_pppp_GR}
\hspace{-.4cm}
\Mcal^{(2)}_{4}{=} {-} i\, & \left(\frac{\kappa}{2}\right)^6 \, L^{(2)} (\epsilon)  \frac{\Tcal^2}{240 \epsilon}
\Bigg[
(4 -2N_{\!f }) \, 
(s^2{+}t^2{+}u^2)\, 
\Sigma_{\log}
+ \left( \frac{6862}{45} 
- \frac{N_{\! f} (32744 + 49 N_{\! f})}{1440}
\right) 
s t u
\Bigg]
{+} \Ocal(\epsilon^0)\,. 
\hspace{-.6cm}
\end{align}
The $\Ocal(\epsilon^0)$ terms are provided in the ancillary file~\cite{AncillaryFile}.
%
The recurring sum of logarithms 
\begin{equation}
\Sigma_{\log} = s \log \left(\frac{-s}{\mu^2}\right) 
+ t \log\left(\frac{-t}{\mu^2}\right) 
+ u \log\left(\frac{-u}{\mu^2}\right) ,
\label{eq:SigmaDef}
\end{equation}
can be analytically continued to physical $s$-channel scattering kinematics ($s>0$ and $t,u<0$) by setting $\log(\frac{-s}{\mu^2}) \mapsto \log(\frac{s}{\mu^2}) - i \pi$.  The $\Ocal(\epsilon^0)$ terms are too lengthy to print here and are included in the ancillary file~\cite{AncillaryFile}.
Interestingly, each term in the result contains at most three momentum-dependent logarithmic factors.
The reduced transcendentality compared to the typical transcendentality-4 of two-loop amplitudes is a consequence of the vanishing all-plus helicity tree amplitude. In the HV scheme, our pure-gravity results agree with the ancillary results of Ref.~\cite{Abreu:2020lyk}, showing we are on the right track.

Setting $N_{\!f}=2$ cancels the $\Sigma_{\log}$ terms, which are tied to the infrared singularities of the bare amplitude. This cancellation occurs simultaneously with that of the one-loop anomaly. The remaining $\Ocal(\epsilon^{-1})$ pole is proportional to $stu$ and, as we will see, has precisely the structure of the $R^3$ tree-level counterterm amplitude that removes the two-loop UV divergence.

\subsection{Infrared subtraction}
\label{subsec:review_amp_ir_divs}

Loop-level gravitational scattering amplitudes generally have infrared soft singularities~\cite{Weinberg:1965nx}; in dimensional regularization these appear as poles in $\epsilon$. The infrared structure of gravity is much simpler than that of gauge theory due to the absence of collinear singularities; it further simplifies for identical-helicity amplitudes because such amplitudes vanish identically at tree-level. We describe here the subtraction of IR divergences for such amplitudes.  

To all orders, the infrared singularities exponentiate~\cite{Weinberg:1965nx, Naculich:2011ry, Akhoury:2011kq}.  Explicitly at four points, 
\begin{align}
\Hcal_4 = e^{-{\rm \Gcal_4}} \, \Mcal_4\,.
\label{HardPart}
\end{align}
The four-point amplitude $\Mcal_4$ and the IR-finite ``hard'' function $\Hcal_4$ have perturbative expansions in the gravitational coupling $\kappa$.  In our IR prescription, we take the soft factor $\Gcal_4$ to be exact in dimensional regularization,
\begin{align}
\label{eq:gravity_IR_curlG_factor}
\Gcal_4\ =\
\left(\frac{\kappa}{2}\right)^2 
L^{(1)}(\epsilon) \,
\frac{2\mu^2}{\epsilon^2} \, \biggl[
 \left(\frac{-s}{\mu^2}\right)^{1-\epsilon}
+\left(\frac{-t}{\mu^2}\right)^{1-\epsilon}
+\left(\frac{-u}{\mu^2}\right)^{1-\epsilon}\,
\biggr] \,.
\end{align}
Here, the leading $1/\epsilon^2$ divergence is reduced to a $1/\epsilon$ due to $s+t+u=0$; the coefficient of the surviving $1/\epsilon$ is the combination $\Sigma_\text{log}$ of momentum invariants in Eq.~\eqref{eq:SigmaDef}. For the finite two-loop hard function, it is important to retain this expression through $\Ocal(\epsilon^0)$; otherwise, the hard function ${\cal H}_4$ is polluted by terms determined by infrared physics. 
 
Using the fact that the all-plus helicity tree amplitude, $\Mcal^{(0)}_4$, vanishes, we have  a simplified infrared structure, through two-loop order,
\begin{align}
\label{eq:2loop_IR_subtraction}
\begin{split}
\Hcal^{(0)}_4 = 0\,,
\qquad 
\Hcal^{(1)}_4 = \Mcal^{(1)}_4,
\qquad 
\Hcal^{(2)}_4 = \Mcal^{(2)}_4\! - \Gcal_4 {\times} \Mcal^{(1)}_4 \ .
\end{split}
\end{align}
As mentioned, we retain the full $\epsilon$ dependence in $\Gcal_4$, and thus avoid spurious logarithmic terms that may be induced in the finite parts of $\Hcal^{(i)}_4$.
In our prescription, the infrared subtraction term, including its finite contribution, is defined by
\begin{equation}
\Mcal^{(2)}_4(1^+, 2^+, 3^+, 4^+) \Bigr|_{\mathrm{IR\,sub}} \equiv \Gcal_4 \Mcal^{(1)}_4(1^+, 2^+, 3^+, 4^+) \,.
\end{equation}
We use the unexpanded expression for $\Gcal_4$ to specify our finite infrared-subtraction prescription. To determine the two-loop hard function through $\Ocal(\epsilon^0)$, it suffices to retain $\Gcal_4$ through $\Ocal(\epsilon^0)$ and the one-loop amplitude through $\Ocal(\epsilon)$.

To see the importance of this prescription, consider the case when the self-dual anomaly vanishes, and thus the one-loop all-plus amplitude vanishes at $\Ocal(\epsilon^0)$, independently of the regularization scheme. The hard function $\Hcal^{(2)}_4$ then depends on the scheme-dependent $\Ocal(\epsilon)$ terms of $\Mcal^{(1)}_4$, which are promoted to $\Ocal(\epsilon^0)$ by the $1/\epsilon$ pole of the soft factor $\Gcal_4$. Thus, even when $\Mcal^{(2)}_4$ is IR finite, as in \GRmatterPlus, the prescription defined by Eq.~\eqref{eq:gravity_IR_curlG_factor} subtracts finite terms involving IR logarithms.

In the HV regularization scheme used here, bosonic and fermionic states are continued differently outside four dimensions ($D_s=4-2\epsilon$ for bosons versus strictly 4 dimensions for fermions), so the one-loop cancellation at $\Ocal(\epsilon^0)$ need not extend to $\Ocal(\epsilon)$. In principle, one could adjust the regularization scheme so that the number of bosonic states matches the number of fermionic ones even outside four dimensions.\footnote{\linespread{0.9}\selectfont Siegel's dimensional reduction scheme~\cite{Siegel:1979wq} has this property.} The finite subtraction above accounts for this residual contribution within our chosen hard-function prescription.\footnote{\linespread{0.9}\selectfont For an infrared-safe cross section, virtual soft singularities cancel against real-emission contributions. Here we compute only the virtual amplitude and specify its finite infrared subtraction through the definition of the hard function.}

Gravity's simple infrared structure allows a clean subtraction of IR divergences, as included in the assembly formula~\eqref{eq:2-loop_remainder_pppp}. We next discuss certain subtle aspects of UV renormalization in gravity, relying on Refs.~\cite{Bern:2015xsa, Bern:2017puu}.

\subsection{UV renormalization and evanescent effects}
\label{subsec:review_amp_uv_divs}

For UV renormalization, we use the $\overline{\rm MS}$ subtraction scheme, together with the HV regularization prescription specified above. This choice lets us compare the numerical values of the divergences directly with the literature~\cite{Goroff:1985sz, Goroff:1985th, vandeVen:1991gw, Bern:2015xsa, Abreu:2020lyk}. Switching regularization schemes can change these numerical values and introduce additional subtleties.

The long and nuanced history of UV properties of IR-subtracted gravitational amplitudes is intimately tied~\cite{Bern:2015xsa, Bern:2017puu} to evanescent effects, visible only outside strictly $D=4$. Here these arise as counterterm amplitudes involving the Gauss--Bonnet operator, whose perturbative pure-graviton matrix elements vanish in strictly four dimensions, but whose insertions can contribute in dimensionally regulated loop amplitudes or as double insertions at tree level. 

At one loop, pure-gravity amplitudes are UV finite~\cite{tHooft:1974toh} in four dimensions.  After accounting for field redefinitions, the only independent on-shell curvature-squared counterterm is the Gauss--Bonnet combination or Euler density,
\begin{equation}
E_4 = R^2-4R_{\mu\nu}^2+R_{\mu\nu\rho\sigma}^2 \,.
\end{equation}
In exactly four dimensions, the density $\sqrt{-g}\, E_4$ is locally a total derivative. Its integral is topological up to boundary terms, so the integrated Gauss--Bonnet term has vanishing perturbative matrix elements about flat space with the usual scattering boundary conditions.  

This four-dimensional finiteness does not imply that the corresponding dimensionally regulated UV divergence vanishes. 
The Gauss--Bonnet operator is evanescent in this sense. In $D=4-2\epsilon$ dimensions, $\sqrt{-g}\, E_4$ is no longer a total derivative, and the dimensionally regulated theory has a nonzero $1/\epsilon$ counterterm coefficient.

Collecting the results from Refs.~\cite{Capper:1974ic, Capper:1975ig, Tsao:1977tj, Gibbons:1978ac, Critchley:1978kb, Duff:1980qv}, and following Ref.~\cite{Bern:2017puu},
the $\overline{\rm MS}$ counterterm is
\begin{equation}
\mathcal L_{\rm GB}^{\rm CT}
 = L^{(1)}(\epsilon)\frac{c_{\rm GB}}{\epsilon}  E_4  \,,
\label{eq:GaussBonnetCounterterm}
\end{equation}
where
\begin{equation}
c_{\rm GB} =  \biggl( 
    \frac{53}{90}  
    + \frac{7 n_{1/2}}{1440} 
    + \frac{n_0}{360} 
    + \frac{91 n_2}{360} 
    - \frac{n_3}{2}
    - \frac{233 n_{3/2}}{1440}  
 \biggr)\, .
 \label{eq:cGBcoefficient}
\end{equation}
Here $n_{1/2}$ is the number of spin-$1/2$ chiral fermions, $n_0$ is the number of real scalars, $n_2$ is the number of two-form fields, $n_3$ is the number of three-form fields, and $n_{3/2}$ is the number of spin-$3/2$ gravitinos.  We define the action according to $S=\int d^D x\,\sqrt{-g}\,\Lcal$.

To obtain finite amplitudes at two loops, we must account for several types of contributions. The (a) bare amplitude, (b) single Gauss--Bonnet counterterm, (c) double Gauss--Bonnet counterterm, and (d) $R^3$ counterterm are illustrated in \fig{fig:TwoloopCounterterm} via representative diagrams. In addition, diagrams with up to five-point interactions must be included. The Gauss-Bonnet counterterm appears in two classes of diagrams.  The first class (b) consists of one-loop diagrams with a single insertion of the Gauss--Bonnet term, and the second class (c) consists of tree diagrams with Gauss--Bonnet insertions.  

After including the Gauss--Bonnet counterterm contributions to remove UV subdivergences, the leftover UV divergence needs to be absorbed by an $R^3$ counterterm, 
\begin{align}
\Lcal_{{R^3}}^{\rm CT} = \biggl(\frac{\kappa}{2} \biggr)^2
L^{(2)} (\eps) \, \frac{c_{{\rm R}^3}}{\epsilon}
R^{\alpha \beta}_{\!\hphantom{\alpha \beta}\gamma \delta}\,
    R^{\gamma \delta}_{\!\hphantom{\gamma \delta}\rho \sigma}\,
    R^{\rho \sigma}_{\!\hphantom{\rho \sigma}\alpha \beta}  \,.
 \label{eq:R3CountertermFourPoint} 
\end{align}
We discussed the second independent $R^3$ operator and the counterterms required for external-matter amplitudes in \sect{subsec:chiral_fermion_theory}. Although the two $R^3$ operators are equivalent for the four-dimensional external-graviton amplitudes considered here, they can lead to distinct subdivergences at three loops.  That is, the two $R^3$ operators differ by an evanescent operator, and thus the three-loop amplitudes exhibit subtleties analogous to those discussed here.

For the graviton and fermion content considered here, the counterterm coefficients are fixed as follows. The $c_{{\rm GB}}$ coefficient of the Gauss-Bonnet counterterm is fixed by one-loop renormalization in $D=4-2\eps$.  In the $\overline{\rm MS}$ scheme it cancels the pure pole piece, giving
\begin{align}
 c_{{\rm GB}} = 
 \left(
    \frac{53}{90} +\frac{7 N_{\!f}}{2880}
 \right)
  .
\label{eq:fourpoint_GB_coefficient}
\end{align}
The divergent part of the $R^3$ counterterm coefficient is fixed to remove the remaining two-loop UV divergence in the $\overline {\rm MS}$ scheme, 
\begin{align}
 c_{\rm R^3} = 
 \left(
    \frac{209}{1440} +  \frac{19 N_{\! f}}{11520}
 \right) + \epsilon \, c_{\rm R^3}^{\rm fin} \equiv c_{\rm R^3, div} + \epsilon \, c_{\rm R^3}^{\rm fin}\, .
\label{R3Divergence}
\end{align}
We also include a finite counterterm $c_{\rm R^3}^{\rm fin}$, to be adjusted to enforce desired properties of the two-loop amplitude.   

The value of the counterterm depends on the specific choices made in defining the HV scheme.  One can adjust the scheme by choosing, for example, $D_s=4$~\cite{Bern:1991aq, Bern:2002zk} to cancel the one-loop anomaly exactly for $N_{\!f} =2$, but this scheme still requires an infinite counterterm~\cite{FuturePaper} and offers no particular advantage over the HV scheme. As emphasized in Ref.~\cite{Bern:2015xsa}, the UV divergence depends on evanescent effects and is scheme dependent, while the renormalization scale dependence is free of these issues.

\subsubsection{Counterterm amplitudes}
\label{subsec:UV_subtraction_explicit}

We compute these via generalized unitarity, sewing three- or four-point counterterm vertices into one-loop and tree-level diagrams, respectively; see Figs.~\ref{fig:TwoloopCounterterm}(b) and (c). For the one-loop diagram with one insertion of the Gauss-Bonnet counterterm, we obtain: 
\begin{align}
\hspace{-.5cm}
\Mcal^{(1)}_{4,{\rm GB}} = i \frac{c_{{\rm GB}}}{\epsilon} \left(\frac{\kappa}{2}\right)^6 \,  L^{(2)}(\epsilon)\, \Tcal^2 
\left[ 
\Bigl(\frac{26}{15} - \frac{7 N_{\!f} }{60} \Bigr)  stu  
+ \Ocal(\epsilon) 
\right] \!.
\hspace{-.5cm}  
\label{eq:1loopGB}
\end{align}
The $1/\epsilon$ term of this amplitude, controlled by the coefficient $c_{{\rm GB}}$ of the Gauss-Bonnet counterterm, cancels the UV divergence of the one-loop (sub)amplitude with external momenta and polarizations outside $D=4$. The required counterterm amplitude through $\Ocal(\epsilon)$ is given in the ancillary files. 

The double-Gauss-Bonnet tree amplitude, given by the diagram in Fig.~\ref{fig:TwoloopCounterterm}(c), is 
\begin{align}
\Mcal^{(0)}_{4,{\rm GB}^2} = +24i \, \left(\frac{c_{{\rm GB}}}{\epsilon}\right)^2 \left(\frac{\kappa}{2}\right)^6 L^{(2)}(\epsilon) \Tcal^2  \frac{ \eps }{1 - \eps} 
s t u\, .
\label{eq:treeGBGB}
\end{align}
The overall $1/\epsilon^2$ comes from the counterterm coefficient, while the $\epsilon$ numerator factor arises because the only non-zero contribution comes from the extra-dimensional components of the graviton exchanged between the two vertices. 
Last, the tree with an insertion of the two-loop $R^3$ counterterm from Eq.~\eqref{eq:R3CountertermFourPoint} is
\begin{align}
\Mcal^{(0)}_{4,{\rm R}^3} = - 60 i \frac{c_{{\rm R}^3}}{\epsilon} \,\left(\frac{\kappa}{2}\right)^6   L^{(2)}(\epsilon) \Tcal^2\,  
s t u \,. 
\label{eq:treeR3}
\end{align}
See also the discussion below Eq.~\eqref{eq:R3CountertermFourPoint}.


\subsubsection{Evanescent effects and renormalization-scale dependence}

Evanescent counterterms can modify the numerical coefficient of the two-loop ultraviolet pole without changing the renormalization-scale dependence of the renormalized amplitude~\cite{Bern:2015xsa, Abreu:2020lyk}. The latter governs the running of the $R^3$ coupling and is central to interpreting the ultraviolet behavior of the amplitudes considered here.

Interestingly, the coefficient of the divergence is directly related to the conformal or trace anomaly~\cite{Capper:1974ic, Capper:1975ig, Tsao:1977tj, Gibbons:1978ac, Critchley:1978kb, Duff:1980qv}.  In particular, the Gauss--Bonnet contribution to the one-loop divergence also determines the coefficient multiplying the Euler density in the anomalous trace of the on-shell energy-momentum tensor,
\begin{equation}
\bigl\langle T^\mu{}_\mu \bigr\rangle
\propto c_{\rm GB}\,E_4 \,.
\end{equation}
In dimensional regularization, the Weyl rescaling of the dimensionally continued Gauss--Bonnet operator is proportional to $D-4$, which combines with its $1/\epsilon$ UV pole to yield a finite trace anomaly as $D\rightarrow 4$. Equivalently, the trace anomaly is obtained from the same divergence by replacing one external graviton polarization tensor with its trace.

In our calculation here, we need the graviton and spin-$1/2$ contributions to \eqn{eq:GaussBonnetCounterterm}, corresponding to a graviton or chiral fermion circulating in the loop, because the Gauss--Bonnet counterterm is required to subtract subdivergences at two loops. As part of the computation in this section, 
we independently confirm the graviton and $n_{1/2}$ coefficients appearing in \eqn{eq:GaussBonnetCounterterm}. The remaining terms in \eqn{eq:GaussBonnetCounterterm} display several noteworthy features. Although a minimally-coupled scalar field is classically dual to an antisymmetric tensor two-form field in four dimensions, their contributions to the Gauss--Bonnet divergence are not identical. This mismatch led Duff and van Nieuwenhuizen to argue that classically equivalent scalar and antisymmetric-tensor descriptions can be quantum mechanically inequivalent~\cite{Duff:1980qv}. Even more strikingly, a three-form field carries no local propagating degrees of freedom in four dimensions, yet contributes both to the trace anomaly and to the one-loop gravitational divergence in \eqn{eq:GaussBonnetCounterterm}-\eqref{eq:cGBcoefficient}.

Interestingly, the trace anomaly coefficient in \eqn{eq:GaussBonnetCounterterm}-\eqref{eq:cGBcoefficient} is modified by duality transformations or by inclusion of the three-form fields. For the two-loop four-graviton all-plus amplitude, scalar--two-form duality and the addition of three-form fields leave the renormalization-scale dependence unchanged. The remaining differences can be absorbed into the divergent and finite parts of the $R^3$ counterterm, so the renormalized amplitudes agree after matching the $R^3$ coupling~\cite{Bern:2015xsa, Bern:2017puu, Dunbar:2017qxb, Abreu:2020lyk}. The one-loop results feed into two-loop amplitudes in rather interesting ways, affecting both infrared and ultraviolet features.  Because the one-loop divergence is non-vanishing outside four dimensions, dimensional regularization requires including the corresponding counterterm \eqref{eq:GaussBonnetCounterterm} to subtract UV subdivergences.   In these subdiagrams, the dimensionally regulated momenta and state sums probe directions outside four dimensions, so the four-dimensional topological argument does not eliminate the Gauss--Bonnet contribution. 
There are also other ways to account for the Gauss--Bonnet subdivergence. In Refs.~\cite{Goroff:1985sz, vandeVen:1991gw}, the subdivergences were subtracted integral by integral rather than by explicitly inserting the GB counterterm, thereby implementing the same subtraction. 
In the subsequent calculations of Refs.~\cite{Bern:2015xsa, Bern:2017puu, Dunbar:2017qxb, Abreu:2020lyk}, the GB counterterm was instead included explicitly, as we do here.  Regardless of how the subdivergence is subtracted, its contribution must be included to obtain the correct expression for the amplitude, and in particular to correctly determine the renormalization-scale dependence. In particular, the GB counterterm in \eqn{eq:GaussBonnetCounterterm} plays a crucial role in the renormalized four-point two-loop all-plus helicity amplitude which we assemble in~\sect{sec:renormalized4PTamplitude}.

Ref.~\cite{Bern:2015xsa} computed the coefficient of the $R^3$ counterterm for minimally coupled theories with a scalar, two-form, and three-form, each of which couples only gravitationally. In the  $\overline {\rm MS}$ scheme~it~is 
\begin{equation}
c_{{\rm R}^3,{\rm div}} = \frac{209}{1440} - \frac{1}{2880} n_0 + \frac{299}{2880} n_2  - \frac{1}{8} n_3 \, . 
\label{eqn:R3CoeffPrevious}
\end{equation}
The first term above was first computed by Goroff and Sagnotti~\cite{Goroff:1985sz, Goroff:1985th}, and verified in multiple calculations~\cite{vandeVen:1991gw, Bern:2015xsa, Abreu:2020lyk}, including the current paper. For the chiral theory considered here, the fermion contribution to the $c_{{\rm R}^3}$ coefficient is given in Eq.~\eqref{R3Divergence}.  

After subtracting the infrared contribution and all one-loop ultraviolet subdivergences, the remaining ultraviolet pole and explicit renormalization-scale dependence of the four-point all-plus amplitude take the form
\begin{align}
  \mathcal{M}_{4}^{(2),\rm sub} (1^+,2^+,3^+,4^+) = 
  i \left(\frac{\kappa}{2} \right)^6 L^{(2)} (\epsilon) \,  \mathcal{T}^2\, stu 
  \biggl( 
    \frac{60\, {c_{{\rm R}^3,{\rm div}}}}{\epsilon} - \frac{N_{b} - N_{\! f}}{8} \log\mu^2 
  \biggr)  + \text{finite}\, .
   \label{AmplUV}
\end{align}
Here ``sub'' denotes subtraction of infrared contributions and ultraviolet subdivergences, with the overall $R^3$ counterterm still to be included. For the graviton--fermion theory considered here,
\begin{equation*}
\mathcal{M}_{4}^{(2),\rm sub}
= \mathcal{M}_{4}^{(2)} - \Gcal_4\mathcal{M}_{4}^{(1)}
+ \mathcal{M}_{4,\rm GB}^{(1)} + \mathcal{M}_{4,{\rm GB}^2}^{(0)}\,.
\end{equation*}
For other matter content, any additional required subdivergence counterterms are included in this definition.

The $N_{b} - N_{\! f}$ coefficient in front of the renormalization scale dependence $\log\mu^2$ 
is the difference between the number of four-dimensional bosonic and fermionic physical states.  For the coefficient in \eqn{eqn:R3CoeffPrevious}, we have
\begin{equation}
N_{b} = 2 + n_0 + n_2 \,, \hskip 1.5 cm N_{\! f} = 0 \,.
\
\end{equation}
As in the one-loop amplitude, the factor of $\mathcal{T}^2$ with $\mathcal{T}$ in Eq.~\eqref{eq:Tfactor} accounts for the helicity weight of the amplitude, and $s,t,u$ are the Mandelstam variables. While a three-form field does not carry physical degrees of freedom in $D=4$, it does contribute to the counterterm, but not to the renormalization scale dependence.  

The remarkable feature of \eqn{AmplUV} is the complete disconnect between the coefficient of the $1/\epsilon$ divergence and the violation of the classical scaling behavior of the amplitude induced by renormalization-scale dependence. While the renormalization-scale dependence has direct physical consequences, the numerical coefficient of the two-loop ultraviolet pole is scheme dependent and is absorbed into the $R^3$ counterterm. This pole coefficient alone does not determine the physical scale dependence.  
Finite local terms proportional to the $R^3$ matrix element also depend on the regularization and subtraction prescriptions. Their scheme dependence is compensated by a corresponding finite redefinition of the $R^3$ coupling. Once that coupling is fixed by a renormalization or matching condition, the combined contribution to the amplitude is fixed~\cite{Bern:2015xsa, Bern:2017puu}. 

The disconnect between the coefficient of the divergence and the renormalization-scale dependence arises from the interplay between the bare contribution and the evanescent subdivergences. Exposing the $\mu$ dependence of each of the four types of terms in \fig{fig:TwoloopCounterterm},
\begin{equation}
\frac{C^{\rm (a)}}{\eps} \,(\mu^2)^{2 \eps} + \frac{C^{\rm (b)}}{\eps}\, (\mu^2)^{\eps} 
 +  \frac{C^{\rm (c)}}{\eps} +  \frac{C^{\rm (d)}}{\eps} 
 = \Big(C^{\rm (a)} + C^{\rm (b)} + C^{\rm (c)} + C^{\rm (d)}  \Bigr) \frac{1}{\eps} + 
 (2 C^{\rm (a)} + C^{\rm (b)}) \ln\mu^2 + \cdots \,.
\label{eq:DivergenceLogMu}
\end{equation}
where the $C^{\rm (x)}$ are the coefficients in front of the $1/\epsilon$ terms for each contribution type. The power of $\mu$ is governed by the number of loops of the diagram and follows from dimensional analysis of the corresponding integrals, together with the fact that counterterm insertions themselves do not carry factors of $(\mu^2)^{\eps}$ in the $\overline{\rm MS}$ scheme.
Thus, the coefficient of the $1/\eps$ divergence receives contributions from all four terms, while the $\ln(\mu^2)$ dependence depends only on those carrying explicit powers of $\mu^{2\eps}$. 
The coefficient $C^{\rm (c)}$, in particular, is highly sensitive to the details of the analytic continuation out of four dimensions.

The simple renormalization-scale dependence can be understood directly from four-dimensional unitarity~\cite{Bern:2017puu, Dunbar:2017qxb}. After subtracting the universal infrared contribution, the cuts determine the coefficients of the kinematic logarithms, whose dimensionless arguments then fix the accompanying $\log\mu^2$ dependence. For the all-plus amplitude, the relevant two-particle cuts contain the one-loop all-plus amplitude in Eq.~\eqref{eq:FourPointOneLoopAmplGen}. Its overall factor of $N_b-N_{\!f}$ explains the corresponding dependence in Eq.~\eqref{AmplUV}, despite the more complicated coefficients of the individual ultraviolet divergences.

The assembly in Eq.~\eqref{eq:2-loop_remainder_pppp} can be checked directly using the coefficients in Eq.~\eqref{eq:DivergenceLogMu}. After removing the infrared contribution and factoring out $i(\kappa/2)^6 L^{(2)}(\epsilon)\Tcal^2stu$, let $C^{\rm(a)}$, $C^{\rm(b)}$, $C^{\rm(c)}$, and $C^{\rm(d)}$ denote the local $1/\epsilon$ coefficients of the bare amplitude, the single and double Gauss--Bonnet insertions, and the $R^3$ counterterm, respectively. UV pole cancellation and the renormalization-scale dependence require
\begin{align}
& C^{\rm(a)}+C^{\rm(b)}+C^{\rm(c)}+C^{\rm(d)} = 0\,,
\nonumber \\
& 2C^{\rm(a)}+C^{\rm(b)} = -\frac{N_b-N_{\!f}}{8}=\frac{N_{\!f}-2}{8}\,,
\end{align}
where $N_b=2$ for the theory considered here, where the only bosonic states are those of the graviton. The first relation fixes the divergent part of the $R^3$ counterterm, while the second checks the scale dependence independently. For a single chiral fermion, $N_{\!f}=2$, the scale-dependent combination vanishes. Once the UV poles are canceled, the finite terms in the same assembly give the renormalized amplitude presented next.

\subsection{Renormalized two-loop four-point all-plus helicity amplitudes}
\label{sec:renormalized4PTamplitude}

%
Combining all contributions in Eq.~\eqref{eq:2-loop_remainder_pppp}, with the counterterm amplitudes in Eqs.~\eqref{eq:1loopGB}, \eqref{eq:treeGBGB} and \eqref{eq:treeR3} and counterterm coefficients in Eqs.~\eqref{eq:fourpoint_GB_coefficient} and~\eqref{R3Divergence}, gives the IR-subtracted UV-renormalized two-loop amplitude. As noted above, the divergence itself is scheme-dependent and can be nonzero even in theories that should be treated as two-loop finite, such as supergravity. 
We note that the non-minimal four-fermion coupling $c_{4\Psi}$ drops out from the divergent part of the unsubtracted amplitude and therefore also from the counterterms. Indeed, its contribution factorizes into two one-loop subloops, so all its divergences are inherited from those of the one-loop factors. However, on dimensional grounds no counterterm is available for either factor, so the contribution of the non-minimal four-fermion term must be finite.
The resulting renormalized gravitational amplitude is
\begin{align}
\Mcal^{(2),\rm ren}_{4} = i\, \left(\frac{\kappa}{2}\right)^6 L^{(2)} (0) \, \Tcal^2
 & 
\bigg[
 stu 
  \bigg(\! {-}60 \, c_{{\rm R}^3}^{\rm fin}
    +\frac{117617}{21600} 
    +\frac{75359}{172800}N_{\!f}
    +\frac{(23 - 194400\, c_{4\Psi})}{345600} N^2_{\!f} 
 \bigg) \nonumber\\
& \hskip .3 cm 
+\frac{1}{120} \left(N_{\!f} - 2\right) (f_{stu}+f_{tus}+f_{ust}) 
\bigg]\, . 
\end{align}   
where $f_{stu} = s \left(3t^2 + 3 u^2 -2 s^2\right) \log (-s/\mu^2)$, etc.  Setting $N_{\! f}=2$, corresponding to the anomaly-free chiral theory, cancels all logarithms in $f_{stu}$, $f_{tus}$, and $f_{ust}$. The renormalized two-loop amplitude is therefore rational
\begin{equation}
\Mcal^{(2),\rm ren}_{4,\rm chiral}
= 10\,i\,\alpha\left(\frac{\kappa}{2}\right)^2\Tcal^2\,stu\,,
\label{eq:AmplChiral}
\end{equation}
where the coefficient $\alpha$ depends on the finite counterterm, 
\begin{equation}\alpha = \left(\frac{\kappa}{2}\right)^4  L^{(2)}(0)
\frac{1}{1920}\label{eq:chiral_alpha}
\left(
1213
-11520\,c^{{\rm fin}}_{{\rm R}^3}
-432\,c_{4\Psi}
\right)
\,.
\end{equation}
Its normalization was chosen so Eq.~\eqref{eq:AmplChiral} agrees with the tree amplitude with one $R^3$ insertion, as in Ref.~\cite{Dunbar:2017dgp}. It also agrees with the four-point KLT assembly in Eq.~\eqref{eq:chiral_KLT_fourpoint_match}.
Since the finite counterterm coefficient $c^{{\rm fin}}_{{\rm R}^3}$ can be adjusted\footnote{\linespread{0.9}\selectfont Recall that the inclusion of finite local counterterms is part of the definition of a quantum theory. Such counterterms may be adjusted to, e.g., restore symmetries, cancel certain anomalies, etc. In contrast, the four-fermion contact coefficient $c_{4\Psi}$ in Eq.~\eqref{LagrangianChiral} is part of the definition of the classical theory. One may, of course, adjust it to set $\alpha=0$; this, however, leads to a classically distinct gravitational theory.}, $\alpha$ can take any value, including zero. 
The absence of $\log(\mu^2)$ terms then implies that a vanishing of the four-point all-plus amplitude is a stable renormalization-group flow at two loops.

\section{Conclusions and Outlook}
\label{sec:conclusions}

The identical-helicity amplitudes of massless gauge and gravity theories are remarkably simple: they vanish at tree level and would also vanish at one loop
were it not for an anomaly~\cite {Bardeen:1995gk, Cangemi:1996rx, Doran:2023cmj, Costello:2022upu}. Matter coupling contributes to this anomaly and, in gravitational theories with the number of fermionic states matching the number of bosonic states, this anomaly cancels.

We have obtained the two-loop four-point all-plus graviton amplitude in gravity coupled to a massless chiral fermion by explicitly constructing and integrating its integrand, and bootstrapped it to arbitrary multiplicity.
This four-point amplitude equals the all-plus amplitude with a single $\Sigma\Gamma\Gamma$ vertex computed in self-dual gravity with the anomaly canceled by a chiral fermion, as $\Sigma\Gamma\Gamma$ is the operator that deforms self-dual gravity into general relativity.
After infrared subtraction and ultraviolet renormalization, all logarithmic terms of the four-point amplitude cancel, leaving a purely rational result. This is the gravitational counterpart of the simplification found in gauge theory, where two-loop all-plus amplitudes with suitable anomaly-free matter content have been shown to be purely rational~\cite{Costello:2023vyy, Dixon:2024mzh}.

As a check, we verified that our pure-gravity result reproduces the all-plus helicity amplitude of Ref.~\cite{Abreu:2020lyk}. This calculation was part of a broader effort that treats all helicity configurations simultaneously using formal polarizations; we present further details elsewhere~\cite{FuturePaper}. A key ingredient was the careful treatment of dimensional-regularization subtleties, in particular evanescent effects~\cite{Bern:2015xsa}, which can generate ultraviolet divergences unrelated to the running of physical couplings, and residual finite logarithms, which can arise even in infrared-finite amplitudes.

Up to an overall constant, the anomaly-free four-point two-loop all-plus result is precisely the amplitude generated by a single insertion of the $R^3$ counterterm in Einstein gravity. This relation fixes the normalization of the $n$-point bootstrap construction, as shown in Eq.~\eqref{eq:chiral_KLT_fourpoint_match}. It also implies that the amplitude’s overall normalization is effectively arbitrary, since one may define a quantum theory by extending the classical action with an arbitrary selection of finite local counterterms.

The single-insertion $R^3$ amplitudes can be bootstrapped using on-shell recursion~\cite{Cohen:2010mi}. Ref.~\cite{Dunbar:2017dgp} gives explicit results through seven points; here we obtained a closed-form $n$-point expression by combining KLT relations with a chiral-algebra bootstrap of the corresponding $F^3$ gauge-theory amplitudes, following the framework of Ref.~\cite{Costello:2022wso}. Equation~\eqref{eq:chiral_allplus_KLT} assembles these ingredients with the normalization fixed by the four-point result. As independent checks of the construction, we explicitly verified crossing symmetry and the expected soft limits through 13~points.

Several directions for future work follow naturally.  While we have explicitly demonstrated the rationality of the two-loop four-point identical-helicity amplitude in gravity coupled to a massless chiral fermion and constructed its extension to arbitrary multiplicity, this simplification likely holds for any massless matter content whenever the number of bosonic states balances the fermionic ones.  It would be good to construct a general proof.

Another remarkable feature is that the anomaly is directly intertwined with the UV properties of gravity theories~\cite{Bern:2015xsa, Bern:2017puu, Dunbar:2017qxb}. When the anomaly vanishes, there is no scale dependence associated with the potential $R^3$ counterterm at two loops, so the $n$-graviton amplitudes should be considered UV finite.  Closely related anomalies arise in supergravities with $\mathcal{N}\leq4$~\cite{Marcus:1985yy, Bossard:2012xs, Carrasco:2013ypa, Bern:2013uka, Bern:2017rjw, Bern:2019isl, Kallosh:2026roi} and are intimately intertwined with the encountered four-loop $D^2R^4$ ultraviolet divergence of pure $\mathcal{N}=4$ supergravity~\cite{Bern:2013uka}.  

More generally, evanescent effects can affect even basic properties of supergravity amplitudes, for example, the gauge invariance of amplitudes involving gravitinos. A careful analysis of these issues is particularly important at high loop orders, such as in reexamining the four-loop divergence of $\mathcal{N}=4$ supergravity~\cite{Bern:2013uka} and its connection to the duality anomaly~\cite{Marcus:1985yy}. It is also intriguing that $\Ncal{=}5$ supergravity, which does not possess this anomaly~\cite{Freedman:2017zgq}, exhibits enhanced ultraviolet cancellations~\cite{Bern:2014sna} for which no conventional symmetry explanation is presently known~\cite{Freedman:2018mrv}.  Clearly, much more remains to be understood about the role of the anomaly in the divergences of (super)gravity theories~\cite{Bern:2017rjw, Bern:2019isl, Monteiro:2022nqt}.

Perhaps the most striking consequence of our results is that we can set all two-loop three- and four-point identical-helicity amplitudes in gravity coupled to a chiral fermion to zero by choosing an appropriate finite coefficient for the local $R^3$ counterterm. Because the renormalization-scale dependence vanishes, this coefficient does not run, and the renormalized amplitude vanishes at all scales. Moreover, the recursive structure of the $n$-point two-loop amplitudes extends this cancellation to arbitrary multiplicity, implying that the all-plus sector vanishes at two loops without supersymmetry.  In supersymmetric theories, Ward identities enforce vanishing all-plus amplitudes~\cite {Grisaru:1976vm, Grisaru:1977px}. 
Single-minus amplitudes provide a complementary test, since they also vanish in supersymmetric theories~\cite{Grisaru:1976vm, Grisaru:1977px}; it will be interesting to study their properties for the chiral theory~\cite{FuturePaper}. At higher loop orders, the higher-derivative analogs of $c_{4\Psi}$ may run, so it remains to be determined whether setting the all-plus amplitude to zero can remain a stable renormalization condition. Our results raise the question of what symmetry, if any, singles out this renormalization condition in the absence of supersymmetry and whether the all-plus sector has any protected properties at three loops and beyond.


\subsection*{Acknowledgments}
We especially thank Lance Dixon for his contributions to this project and his help throughout.
We also thank Roland Bittleston and Natalie Paquette for helpful discussions.
Z.B. and E.H.~are supported in part by the U.S. Department of Energy (DOE) under award number DE-SC0009937.
R.R.~is supported by the U.S.~Department of Energy (DOE) under award number~DE-SC00019066 and by a Senior Fellowship at the Institute for Theoretical Studies, ETH Z\"urich.
M.R. is supported by the Department of Energy (DOE), Contract DE-AC02-76SF00515. 
A.M.~is supported by the Simons Foundation as a part of the Simons Collaboration on Celestial Holography.
This research used resources of the National Energy Research Scientific Computing Center (NERSC), a Department of Energy User Facility, using NERSC award ERCAP 0034824.
We are grateful to the Mani L. Bhaumik Institute for Theoretical Physics for support.

\appendix
\section{Computing the all-plus form factor of $F^3$}
\label{app:F3_derivation}

In this appendix, we derive the all-plus tree-level form factor used in Sec.~\ref{subsec:F3_form_factor}, using the chiral algebra bootstrap reviewed in Sec.~\ref{subsec:cab_framework}. The operator $F_+^3$ is the cubic color-trace operator built from the self-dual field strength defined there, with spacetime scaling dimension $\Delta(F_+^3)=6$.

We use positive-helicity gluon states $J^a(\tilde\lambda_i,z_i)$ with momentum $p_i^{\dot\alpha\alpha}=\tilde\lambda_i^{\dot\alpha}\lambda_i^\alpha$, choosing $\lambda_i=(z_i,1)$ so that $\spa i.j=z_i-z_j$ and absorbing the energy into $\tilde\lambda_i$. The nonnegative integers $k_i$ label the energy modes in Eq.~\eqref{eq:modeex}: $J^a[k_i](z_i)$ is homogeneous of degree $k_i$ in $\tilde\lambda_i$, whose dependence is suppressed in this notation. We use the mode OPE in Eq.~\eqref{JJmodeope}. The color-ordered correlators below omit the overall factor of $i$ included in $A_n^{F^3}$, with normalization fixed by Eq.~\eqref{eq:F3_normalization}.

Expanding the form factor in energy modes gives
\begin{equation}
    \label{eq:fftoemodes}
    \langle J^{a_1}(\tilde\lambda_1,z_1)\cdots J^{a_n}(\tilde\lambda_n,z_n)|F_+^3\rangle
    =
    \sum_{\substack{k_1,\dotsc,k_n\geq0 \\ k_1+\cdots+k_n=6}}
    \langle J^{a_1}[k_1](z_1)\cdots J^{a_n}[k_n](z_n)|F_+^3\rangle
    \,,
\end{equation}
where the constraint follows from $\Delta(F_+^3)=6$ and the mode scaling dimensions $\Delta(J[k_i])=-k_i$. Since the form factors are tree-level and involve the scattering of gluons only, each correlator can be written as a permutation sum of traces in the fundamental representation of $SU(N_c)$ times a color-ordered correlator that depends on the kinematics only:
\begin{equation}
\langle J^{a_1}[k_1](z_1)\cdots J^{a_n}[k_n](z_n)|F_+^3\rangle
=
\sum_{\sigma\in S_n/\mathbb{Z}_n}\langle J[k_{\sigma_1}](z_{\sigma_1})\cdots J[k_{\sigma_n}](z_{\sigma_n})|F_+^3\rangle
\,\tr(t^{a_{\sigma_1}}\cdots t^{a_{\sigma_n}})
\,.
\end{equation}
We will focus on computing the color-ordered $\langle J[k_1](z_1)\cdots J[k_n](z_n)|F_+^3\rangle$ for all possible values of the $k_i$. The convenience of working with the color-ordered correlators is that the insertion $J[k_i]$ can only have OPEs with neighboring insertions $J[k_{i-1}]$ and $J[k_{i+1}]$.

We can immediately exclude any correlator with $k_i\geq4$. The $SL(2,\mathbb{C})_+$ symmetry requires that all $\tilde\lambda_i$ are contracted among each other. Because $k_1+\cdots+k_n=6$, a choice of $k_i\geq4$ will necessarily require that a $\tilde\lambda_i$ is contracted with itself, and $[ii]=0$.

For a fixed color ordering, the tree-level OPE permits only simple poles when an insertion approaches either neighbor, with residues determined by lower-point correlators. When the chosen insertion is $J[0]$ or $J[1]$, its conformal weight requires the correlator to vanish at infinity. Any two expressions with the same pole residues therefore differ by a polynomial that vanishes at infinity and, hence, are identical. This establishes the uniqueness of the reconstructions used below.

Let us start with two-point correlators. By $SL(2,C)_+$ symmetry, the only allowed correlator is the one with $k_1=k_2=3$:
\begin{equation}
    \label{eq:33}
    \langle J[3](z_1) J[3](z_2) |F_+^3\rangle = 0
    \,.
\end{equation}
It is zero, because $F_+^3$ does not source a 2-point interaction. At 3 points, the possible correlators are
\begin{align}
\label{eq:330}
&\langle J[3](z_1)J[3](z_2)J[0](z_3)|F_+^3\rangle
=0
\,,
\\
\label{eq:321}
&\langle J[3](z_1)J[2](z_2)J[1](z_3)|F_+^3\rangle
=0
\,,
\\
\label{eq:222}
&\langle J[2](z_1)J[2](z_2)J[2](z_3)|F_+^3\rangle
=-\spb1.2\spb2.3\spb3.1
=\frac{s_{12}s_{23}s_{31}}{\spa1.2\spa2.3\spa3.1}
\,,
\end{align}
and permutations of these correlators that distribute the values of the $k_i$ to different points $z_j$. We can confirm that the first two correlators are zero by treating them as functions of $z_3$ and using the OPEs as $z_3\to z_1,z_2$. Both correlators have zeros at $z_3=\infty$, since $J[0]$ and $J[1]$ have conformal weights 1 and 1/2, respectively, so they are completely determined by the residues of the poles of $z_3$. From the OPE \eqref{JJmodeope}, we see that the residues will be two-point correlators, all of which are zero. The expression for the correlation function \eqref{eq:222} is fixed by $SL(2,\mathbb{C})_+$ symmetry and $z_i=\infty$ behavior, up to an overall coefficient. The coefficient is determined by a simple 3-gluon Feynman diagram computation or by matching to known results \cite{Dixon:2004za, Broedel:2012rc}.

Higher-point correlation functions with a $J[3]$ insertion will continue to vanish. This vanishing follows by induction on the number of gluons. At higher points, one can always pick a $z_i$ that is the argument of either a $J[0]$ or a $J[1]$. The correlators will always have zeros at $z_i=\infty$, allowing them to be determined by the residues of the $z_i$ poles. These residues will be lower-point correlators involving a $J[3]$ or $J[4]$. Since any correlator with a $J[4]$ and the two-point function \eqref{eq:33} are zero, all higher-point ones must vanish if they have a $J[3]$, by induction.

The remaining calculation has two steps: determine the nonzero seeds with three, four, five, or six nonzero-mode insertions, and then extend each seed to arbitrary multiplicity by inserting $J[0]$ modes. We first explain this second step for the three-point seed~\eqref{eq:222}; the same insertion argument will apply to the higher-point seeds below. We now prove that
\begin{multline}
\label{eq:222npt}
\langle J[0](z_1)\cdots J[2](z_i)\cdots J[2](z_j)\cdots J[2](z_k)\cdots J[0](z_n)|F_+^3\rangle
=\frac{s_{ij}s_{jk}s_{ki}}{\spa1.2\spa2.3\cdots\spa n.1}
\\
\equiv
\frac{T_3(i,j,k)}{\spa1.2\spa2.3\cdots\spa n.1}
\,,
\end{multline}
where the ellipses are insertions of $J[0]$. For the induction step, $n>3$, cyclically relabel the ordering so that leg $n$ is a $J[0]$ insertion; the three $J[2]$ insertions then have indices $1\leq i<j<k\leq n-1$. The correlator has a zero of order 2 at $z_n=\infty$, so it is determined by its residues at poles of $z_n$. The residues are determined by the OPEs as $z_n\to z_{n-1}$ and $z_n\to z_1$. We have
\begin{multline}
\langle J[0](z_1)\cdots J[2](z_i)\cdots J[2](z_j)\cdots J[2](z_k)\cdots J[0](z_n)|F_+^3\rangle
\\
\overset{z_n\to z_{n-1}}{\sim}
\frac{1}{\spa{n-1,}.n}\langle J[0](z_1)\cdots J[2](z_i)\cdots J[2](z_j)\cdots J[2](z_k)\cdots J[0](z_{n-1})|F_+^3\rangle
\end{multline}
and
\begin{multline}
\langle J[0](z_1)\cdots J[2](z_i)\cdots J[2](z_j)\cdots J[2](z_k)\cdots J[0](z_n)|F_+^3\rangle
\\
\overset{z_n\to z_1}{\sim}
\frac{1}{\spa n.1}\langle J[0](z_1)\cdots J[2](z_i)\cdots J[2](z_j)\cdots J[2](z_k)\cdots J[0](z_{n-1})|F_+^3\rangle
\,.
\end{multline}
So, 
\begin{multline}
\langle J[0](z_1)\cdots J[2](z_i)\cdots J[2](z_j)\cdots J[2](z_k)\cdots J[0](z_n)|F_+^3\rangle
\\
=
\left(\frac{1}{\spa{n-1,}.n}+\frac{1}{\spa n.1}\right)
\langle J[0](z_1)\cdots J[2](z_i)\cdots J[2](z_j)\cdots J[2](z_k)\cdots J[0](z_{n-1})|F_+^3\rangle
\\
=
\left(\frac{\spa{n-1,}.n + \spa n.1}
{\spa{n-1,}.n\spa n.1}\right)
\frac{s_{ij}s_{jk}s_{ki}}{\spa1.2\spa2.3\cdots\spa{n-1,}.1}
=
\frac{s_{ij}s_{jk}s_{ki}}{\spa1.2\spa2.3\cdots\spa{n}.1}
\,,
\end{multline}
where we used
\begin{equation}
\spa{n-1,}.n + \spa n.1 = (z_{n-1}-z_n) + (z_n-z_1) = z_{n-1}-z_1=\spa{n-1,}.1
\end{equation}
to cancel the factor of $\spa{n-1,}.1$ in the denominator. Note that the computation above works for $i\geq 1$ and $k\leq n-1$, since the OPE of a $J[0]$ with a $J[m]$ is
\begin{equation}
J[0](z)J[m](w)\sim \frac{1}{z-w}J[m](w)
\,.
\end{equation}
This OPE leaves the numerator $s_{ij}s_{jk}s_{ki}$ unchanged. The sum of the two neighboring pole contributions replaces the denominator factor $\spa{n-1,}.1$ by $\spa{n-1,}.n\spa n.1$, thereby inserting leg $n$ into the cyclic Parke--Taylor denominator. Since the $n=3$ seed~\eqref{eq:222} agrees with Eq.~\eqref{eq:222npt}, repeated $J[0]$ insertions establish the result for arbitrary $n$. This is the insertion step used to extend each of the higher-point seeds below.

At 4 points, the only new correlation functions are
\begin{align}
\label{eq:2211}
&\langle J[2](z_1)J[2](z_2)J[1](z_3)J[1](z_4) |F_+^3\rangle
= \frac{s_{12}s_{23}s_{41} + s_{12}\spa2.3\spb3.1\spa1.4\spb4.2}{\spa1.2\spa2.3\spa3.4\spa 4.1}
\,,
\\
\label{eq:2121}
&\langle J[2](z_1)J[1](z_2)J[2](z_3)J[1](z_4) |F_+^3\rangle
= 0 \,,
\end{align}
and cyclic permutations of these that send $z_i\to z_{i+1}$. We compute these two correlators by looking at their OPEs as $z_4\to z_3$ and $z_4\to z_1$. Starting with the correlator \eqref{eq:2211}, we have
\begin{multline}
\langle J[2](z_1)J[2](z_2)J[1](z_3)J[1](z_4) |F_+^3\rangle
\overset{z_4\to z_3}{\sim}
\frac{1}{\spa3.4}\tilde\lambda_4^{\dot\alpha}\frac{\partial}{\partial\tilde\lambda_3^{\dot\alpha}}
\langle J[2](z_1)J[2](z_2)J[2](z_3) |F_+^3\rangle
\\
= -\frac{1}{\spa3.4}\tilde\lambda_4^{\dot\alpha}\frac{\partial}{\partial\tilde\lambda_3^{\dot\alpha}}\spb1.2\spb2.3\spb3.1
=-\frac{\spb1.2\spb2.3\spb4.1+\spb1.2\spb2.4\spb3.1}{\spa3.4}
=\frac{s_{12}s_{23}s_{41} + s_{12}\spa2.3\spb3.1\spa1.4\spb4.2}{\spa1.2\spa2.3\spa3.4\spa4.1}
\,.
\end{multline}
The $z_4\to z_1$ limit produces a $J[3]$, so the residue vanishes. The limits $z_4\to z_1$ and $z_4\to z_3$ of the correlator in eq.~\eqref{eq:2121} vanish for the same reason. We conclude that eqs.~\eqref{eq:2211} and \eqref{eq:2121} hold.

The $n$-point generalizations of the correlators \eqref{eq:2211} and \eqref{eq:2121} are
\begin{align}
&\langle\cdots J[2](z_i)\cdots J[2](z_j)\cdots J[1](z_k)\cdots J[1](z_l)\cdots|F_+^3\rangle
=
\frac{s_{ij}s_{jk}s_{li} + s_{ij}\spa j.k\spb k.i\spa i.l\spb l.j}
{\spa1.2\spa2.3\cdots\spa n.1}
\\
&\langle\cdots J[2](z_i)\cdots J[1](z_j)\cdots J[2](z_k)\cdots J[1](z_l)\cdots|F_+^3\rangle
= 0
\,.
\end{align}
Starting from the corresponding four-point seeds, the $n$-point formulas follow by repeated $J[0]$ insertions using the induction step above: the numerator is unchanged, while each insertion extends the cyclic denominator. We then see that 
\begin{multline}
T_4(i,j,k,l) = \spa1.2\spa2.3\cdots\spa n.1
\langle\cdots J[2](z_i)\cdots J[2](z_j)\cdots J[1](z_k)\cdots J[1](z_l)\cdots|F_+^3\rangle
\\
+ \mathcal{C}(i,j,k,l)
\,.
\end{multline}
The sum over cycles comes from the 3 other cyclic permutations at 4 points that were mentioned earlier.

The new 5-point correlators are
\begin{multline}
\label{eq:21111}
\langle J[2](z_1)J[1](z_2)J[1](z_3)J[1](z_4)
J[1](z_5)|F_+^3\rangle
\\
=\frac{1}{\spa1.2\spa2.3\cdots\spa 5.1}
\big(s_{12}s_{34}s_{51} 
+ s_{12}\spa5.1\spb1.4\spa4.3\spb3.5
\\
+ s_{51}\spa1.2\spb2.4\spa4.3\spb3.1
+ \spa5.1\spb1.4\spa4.3\spb3.1\spa1.2\spb2.5
\big)
\end{multline}
and cyclic permutations $z_i\to z_{i+1}$ of this correlator. The correlator must have a zero of order 1 at $z_5=\infty$, so it is computable from its residues at $z_5=z_4$ and $z_5=z_1$. The OPE as $z_5\to z_1$ produces a $J[3]$ insertion, so it vanishes. Thus, the correlator \eqref{eq:21111} is given by its residue at $z_5=z_4$:
\begin{multline}
\langle J[2](z_1)J[1](z_2)J[1](z_3)J[1](z_4)
J[1](z_5)|F_+^3\rangle
\\
\overset{z_5\to z_4}{\sim}
\frac{1}{\spa4.5}
\tilde\lambda_5^{\dot\alpha}\frac{\partial}{\partial\tilde\lambda_4^{\dot\alpha}}\langle J[2](z_1)J[1](z_2)J[1](z_3)J[2](z_4)|F_+^3\rangle
\\
=
-\frac{1}{\spa4.5}
\tilde\lambda_5^{\dot\alpha}\frac{\partial}{\partial\tilde\lambda_4^{\dot\alpha}}
\frac{\spb4.1\spb1.2\spb3.4 + \spb4.1\spb1.3\spb2.4}{\spa2.3}
\\
=-\frac{1}{\spa2.3\spa4.5}
\big(
\spb4.1\spb1.2\spb3.5 + \spb5.1\spb1.2\spb3.4
+\spb4.1\spb1.3\spb2.5 + \spb5.1\spb1.3\spb2.4
\big)
\,,
\end{multline}
the last line showing that eq.~\eqref{eq:21111} holds.

The $n$-point generalization of eq.~\eqref{eq:21111} is
\begin{multline}
\label{eq:211111npt}
\langle \cdots J[2](z_i)\cdots J[1](z_j)
\cdots J[1](z_k)\cdots J[1](z_l)\cdots
J[1](z_m)\cdots|F_+^3\rangle
\\
= \frac{1}{\spa1.2\spa2.3\cdots\spa n.1}
\big(s_{ij}s_{kl}s_{mi} 
+ s_{ij}\spa m.i\spb i.l\spa l.k\spb k.m
\\
+ s_{mi}\spa i.j\spb j.l\spa l.k\spb k.i
+ \spa m.i\spb i.l\spa l.k\spb k.i\spa i.j\spb j.m
\big)
\,.
\end{multline}
Starting from the five-point seed~\eqref{eq:21111}, the $n$-point formula follows by the same $J[0]$ insertion step, leaving its numerator unchanged and extending the cyclic denominator. We then see that 
\begin{multline}
T_5(i,j,k,l,m)
\\
= \spa1.2\spa2.3\cdots\spa n.1
\langle \cdots J[2](z_i)\cdots J[1](z_j)
\cdots J[1](z_k)\cdots J[1](z_l)\cdots
J[1](z_m)\cdots|F_+^3\rangle
\\
+ \mathcal{C}(i,j,k,l,m)
\,,
\end{multline}
where the sum over cycles comes from the 4 other 5-point correlators that are given by cyclic permutations of eq.~\eqref{eq:21111}.

There is exactly one 6-point correlation function that is new, and it is
\begin{multline}
\label{eq:111111}
\langle J[1](z_1)\cdots J[1](z_6)|F_+^3\rangle
=-\frac{\spa{1}.{2}\spa{3}.{4}\spa{5}.{6}}{\spa1.2\spa2.3\cdots\spa6.1}
\big(\spb{1}.{2}\spb{3}.{4}\spb{5}.{6} 
+ \spb{6}.{2}\spb{3}.{4}\spb{5}.{1}
+ \spb{1}.{3}\spb{2}.{4}\spb{5}.{6}
\\
+ \spb{1}.{2}\spb{3}.{5}\spb{4}.{6}
+ \spb{6}.{3}\spb{2}.{4}\spb{5}.{1} 
+ \spb{6}.{2}\spb{3}.{5}\spb{4}.{1}
+ \spb{1}.{3}\spb{2}.{5}\spb{4}.{6} 
+ \spb{6}.{3}\spb{2}.{5}\spb{4}.{1}
\big)
+ (123456)
\,,
\end{multline}
where the $+(123456)$ instructs one to add to the expression the single cyclic permutation $1\to2\to\cdots \to6\to1$. Taking $z_6\to z_5$, we have that
\begin{multline}
\langle J[1](z_1)\cdots J[1](z_6)|F_+^3\rangle
\overset{z_6\to z_5}{\sim}
\frac{1}{\spa5.6}\tilde\lambda_6^{\dot\alpha}\frac{\partial}{\partial\tilde\lambda_5^{\dot\alpha}}
\langle J[1](z_1)\cdots J[1](z_4)J[2](z_5)|F_+^3\rangle
\\
=
-\frac{1}{\spa1.2\spa3.4\spa5.6}\tilde\lambda_6^{\dot\alpha}\frac{\partial}{\partial\tilde\lambda_5^{\dot\alpha}}
\big(
\spb3.5\spb5.1\spb2.4 + \spb4.5\spb5.1\spb2.3
+\spb3.5\spb5.2\spb1.4 + \spb4.5\spb5.2\spb1.3\big)
\\
=
-\frac{\spa2.3\spa4.5\spa6.1}{\spa1.2\spa2.3\cdots\spa6.1}
\big(
\spb2.3\spb4.5\spb6.1 + \spb1.3\spb4.5\spb6.2 + \spb2.4\spb3.5\spb6.1 + \spb2.3\spb4.6\spb5.1
\\
+ \spb1.4\spb3.5\spb6.2 + \spb1.3\spb4.6\spb5.2 + \spb2.4\spb3.6\spb5.1 + \spb1.4\spb3.6\spb5.2 
\big)
\,.
\end{multline}
This expression is the $+(123456)$ term in eq.~\eqref{eq:111111}\,. Similarly, the $z_6\to z_1$ limit gives
\begin{multline}
\langle J[1](z_1)\cdots J[1](z_6)|F_+^3\rangle
\overset{z_6\to z_1}{\sim}-\frac{\spa{1}.{2}\spa{3}.{4}\spa{5}.{6}}{\spa1.2\spa2.3\cdots\spa6.1}
\big(\spb{1}.{2}\spb{3}.{4}\spb{5}.{6} 
+ \spb{6}.{2}\spb{3}.{4}\spb{5}.{1}
+ \spb{1}.{3}\spb{2}.{4}\spb{5}.{6}
\\
+ \spb{1}.{2}\spb{3}.{5}\spb{4}.{6}
+ \spb{6}.{3}\spb{2}.{4}\spb{5}.{1} 
+ \spb{6}.{2}\spb{3}.{5}\spb{4}.{1}
+ \spb{1}.{3}\spb{2}.{5}\spb{4}.{6} 
+ \spb{6}.{3}\spb{2}.{5}\spb{4}.{1}
\big)
\,,
\end{multline}
which is the term preceding the $+(123456)$. Since the correlator must have a zero of order 1 at $z_6=\infty$, we conclude that it is given by the sum of these two limits, recovering eq.~\eqref{eq:111111}. The $n$-point generalization is
\begin{equation}
\langle \cdots J[1](z_{i_1})\cdots J[1](z_{i_2})\cdots J[1](z_{i_6})\cdots|F_+^3\rangle
= \frac{T_6(i_1,\cdots i_6)}{\spa1.2\spa2.3\cdots\spa n.1}
\,,
\end{equation}
This extends the six-point seed~\eqref{eq:111111} by repeated $J[0]$ insertions, using the same induction step as for Eq.~\eqref{eq:222npt}.

We have exhausted the list of possible correlation functions that can appear in the mode sum~\eqref{eq:fftoemodes}. Summing over the positions of the three $J[2]$ insertions gives the $T_3$ contributions in Eq.~\eqref{eq:T3}. The cyclic placements of the four- and five-insertion seeds give the sums defining $T_4$ and $T_5$ in Eqs.~\eqref{eq:T4} and~\eqref{eq:T5}, while the six $J[1]$ insertions give $T_6$ in Eq.~\eqref{eq:T6}. Additional $J[0]$ insertions supply the full cyclic denominator without changing these numerators. Restoring the overall factor of $i$ fixed by Eq.~\eqref{eq:F3_normalization}, and summing over all ordered subsets of nonzero-mode positions, therefore reproduces the all-plus form factor $A_n^{F^3}\equiv A_n^{F_+^3}$ in Eq.~\eqref{eq:F3}.

The derivation does not impose momentum conservation on the external gluons: the operator supplies their total momentum $P=\sum_{i=1}^n p_i$, which need not vanish. The momentum-conserving amplitude with a single $F^3$ insertion is obtained by subsequently setting $P=0$, as described in Sec.~\ref{subsec:F3_form_factor}, and is the gauge-theory input to the KLT construction.

\bibliographystyle{apsrev4-1}

\bibliography{GravityTwoLoopAllPlus.bib}

\pagebreak

\end{document}